\documentclass[lettersize,journal]{IEEEtran}
\usepackage[T1]{fontenc}
\usepackage{amsmath,amsfonts}
\usepackage{algorithmic}
\usepackage{array}
\usepackage[caption=false,font=normalsize,labelfont=sf,textfont=sf]{subfig}
\usepackage{textcomp}

\usepackage{stfloats}
\usepackage{verbatim}
\usepackage{graphicx}
\usepackage{hyperref}
\def\BibTeX{{\rm B\kern-.05em{\sc i\kern-.025em b}\kern-.08em
    T\kern-.1667em\lower.7ex\hbox{E}\kern-.125emX}}
\usepackage{balance}

\usepackage{parcolumns}
\usepackage[colorinlistoftodos,textwidth=15mm,textsize=tiny,obeyFinal]{todonotes}
\usepackage{mathtools} 
\usepackage{color}
\usepackage{etoolbox}
\usepackage{colortbl}
\usepackage{framed}
\usepackage{tabularx}
\usepackage[linesnumbered,ruled, vlined]{algorithm2e}
\usepackage{multirow,bigdelim}
\usepackage{enumitem}
\usepackage[utf8]{inputenc}
\usepackage{booktabs}
\usepackage{threeparttable}
\usepackage{adjustbox}
\usepackage{xcolor}
\usepackage{listings}
\usepackage{listings-rust}
\usepackage{tcolorbox}
\tcbuselibrary{skins, breakable}
\usepackage{url}
\usepackage{tikz}
\usetikzlibrary{decorations,calligraphy}
\usepackage{fancyvrb}
\tcbuselibrary{breakable} 
\usepackage{pifont}

\usepackage[english]{babel}
\usepackage{amsthm}

\usepackage[frozencache,cachedir=minted-cache]{minted}
\setminted{
  linenos,
  breaklines,
  autogobble,
  fontsize=\small,
  tabsize=4
}

\usepackage{makecell}

\begin{document}

\title{Evaluating and Improving Static Application Security Testing Tools for Rust}

\author{Huaien Zhang,
        Zhengjie Zhao,
        Yu Pei,
        Tian Dong,
        Minxue Pan,
        Shin Hwei Tan,
        Hao Chen,~\IEEEmembership{Fellow,~IEEE}
\thanks{Huaien Zhang, Tian Dong, and Hao Chen are with the Department of Computer Science, University of Hong Kong, Pok Fu Lam, Hong Kong (e-mail: huaien@hku.hk; tiandong@hku.hk; chenho@hku.hk)}
\thanks{Zhengjie Zhao and Minxue Pan are with the Software Institute, Nanjing University, Nanjing 210023, China (e-mail: zhengjiezhao@smail.nju.edu.cn; mxp@nju.edu.cn).}
\thanks{Yu Pei is with the Department of Computing, Hong Kong Polytechnic University, Hung Hom, Hong Kong (e-mail: max.yu.pei@polyu.edu.hk)}
\thanks{Shin Hwei Tan is with the Department of Computer Science and Software Engineering, Concordia University, Montreal, Canada (e-mail: shinhwei.tan@concordia.ca)}
}

\markboth{IEEE TRANSACTIONS ON SOFTWARE ENGINEERING, OCTOBER~2026}{Zhang \MakeLowercase{\textit{et al.}}: Evaluating and Improving SAST Tools for Rust}

\maketitle

\definecolor{main-color}{rgb}{0.6627, 0.7176, 0.7764}
\definecolor{string-color}{rgb}{0.3333, 0.5254, 0.345}
\definecolor{key-color}{rgb}{0.8, 0.47, 0.196}


\newtcolorbox{systemprompt}{
    width=\textwidth,        
    colback=red!5!white,
    colframe=red!70!black,
    title=System Prompt,
    enhanced,
    halign=left,            
    valign=top,             
    before upper={\parindent0pt\raggedright}, 
}

\newtcolorbox{userpromptwithtitle}{
    width=\textwidth,        
    colback=green!5!white,
    colframe=green!70!black,
    title=User Prompt,
    fonttitle=\bfseries,
    enhanced,
    halign=left,            
    halign title=left,    
    valign=top,             
    before upper={\parindent0pt\raggedright}, 
}

\newtcolorbox{userprompt}{
    width=\textwidth,        
    colback=green!5!white,
    colframe=green!70!black,
    enhanced,
    halign=left,            
    valign=top,             
    before upper={\parindent0pt\raggedright}, 
}


\newtcolorbox{userprompt2}{
    enhanced, 
    width=\textwidth, 
    arc=0pt, 
    outer arc=0pt,
    boxrule=0.5pt, 
    colback=green!5!white, 
    colframe=green!70!black, 
    colbacktitle=green!70!black, 
    coltitle=white, 
    fonttitle=\bfseries\sffamily, 
    title=User Prompt, 
    titlerule=0pt, 
    attach title to upper=
        {\linebreak[0]\vspace{2pt}\rule[0pt]{\linewidth}{0.4pt}\vspace{2pt}},
    breakable,
    left=6pt, 
    right=6pt,
    top=4pt,
    bottom=4pt,
}

\newtcolorbox{aiprompt}{
    colback=purple!5!white,
    colframe=purple!70!black,
    title=AI Response,
    fonttitle=\bfseries,
    breakable,
    enhanced,
}

\newtcolorbox{promptbox}{
    colback=gray!5,      
    colframe=gray!50,     
    arc=3pt,              
    boxsep=5pt,           
    left=6pt, right=6pt, 
    breakable,            
    fontupper=\ttfamily,  
    title={\small\textbf{Prompt}}, 
}

\definecolor{findingblue}{RGB}{46,94,170}
\definecolor{findingbg}{RGB}{247,249,253}

\newcounter{finding}
\renewcommand{\thefinding}{\arabic{finding}}


\newtcolorbox{findingbox}[1][]{
  enhanced,
  breakable,
  colback=findingbg,
  colframe=findingbg,
  boxrule=0pt,
  borderline west={1.5pt}{0pt}{findingblue},
  sharp corners,
  left=2mm,
  right=2mm,
  top=1.2mm,
  bottom=1.2mm,
  before upper={\refstepcounter{finding}\textbf{\color{findingblue}Finding~\thefinding:}\enspace},
  #1
}

\definecolor{GrayCodeBlock}{RGB}{241,241,241}
\definecolor{BlackText}{RGB}{110,107,94}
\definecolor{RedTypename}{RGB}{182,86,17}
\definecolor{GreenString}{RGB}{96,172,57}
\definecolor{PurpleKeyword}{RGB}{184,84,212}
\definecolor{GrayComment}{RGB}{170,170,170}
\definecolor{GoldDocumentation}{RGB}{180,165,45}
\lstdefinelanguage{rust}
{
    columns=fullflexible,
    keepspaces=true,
    frame=single,
    framesep=0pt,
    framerule=0pt,
    framexleftmargin=4pt,
    framexrightmargin=4pt,
    framextopmargin=5pt,
    framexbottommargin=3pt,
    xleftmargin=4pt,
    xrightmargin=4pt,
    backgroundcolor=\color{GrayCodeBlock},
    basicstyle=\ttfamily\color{BlackText},
    keywords={true,false,unsafe,async,await,move,use,pub,crate,super,self,mod,struct,enum,fn,const,static,let,mut,ref,type,impl,dyn,trait,where,as,break,continue,if,else,while,for,loop,match,return,yield,in
    },
    keywordstyle=\color{PurpleKeyword},
    ndkeywords={
        bool,u8,u16,u32,u64,u128,i8,i16,i32,i64,i128,char,str,CString,std,ffi,os,raw,c_char,Self,Option,Some,None,Result,Ok,Err,String,Box,Vec,Rc,Arc,Cell,RefCell,HashMap,BTreeMap,macro_rules
    },
    ndkeywordstyle=\color{RedTypename},
    comment=[l][\color{GrayComment}\slshape]{//},
    morecomment=[s][\color{GrayComment}\slshape]{/*}{*/},
    morecomment=[l][\color{GoldDocumentation}\slshape]{///},
    morecomment=[s][\color{GoldDocumentation}\slshape]{/*!}{*/},
    morecomment=[l][\color{GoldDocumentation}\slshape]{//!},
    morecomment=[s][\color{RedTypename}]{\#![}{]},
    morecomment=[s][\color{RedTypename}]{\#[}{]},
    stringstyle=\color{GreenString},
    string=[b]"
}

\lstdefinestyle{mystyle}{
    language=SQL,
    basicstyle=\ttfamily\footnotesize\color{main-color},
    stringstyle=\color{string-color},
    keywordstyle=\color{key-color},
    keywordstyle=[2]\color{lime},
    keywordstyle=[3]\color{yellow},
    keywordstyle=[4]\color{teal},
    morekeywords=[3]{<<, >>},
    morekeywords=[4]{++},
    basicstyle=\ttfamily\footnotesize,
    commentstyle=\color{blue}\ttfamily,
    morecomment=[f][\lstbg{red!20}]-,
    morecomment=[f][\lstbg{green!20}]+,
    morecomment=[f][\textit]{@@},
    numbers=none,
    texcl=false,
    xleftmargin=0pt,
    xrightmargin=0pt,
    framexleftmargin=0pt,
    framexrightmargin=0pt,
    aboveskip=0pt,
    belowskip=0pt
}

\definecolor{pgrey}{rgb}{0.46,0.45,0.48}

\lstdefinestyle{javastyle}
{
    language = Java,
    basicstyle = {\ttfamily \color{main-color}},
    stringstyle = {\color{string-color}},
    keywordstyle = {\color{key-color}},
    keywordstyle = [2]{\color{lime}},
    keywordstyle = [3]{\color{yellow}},
    keywordstyle = [4]{\color{teal}},
    morekeywords = [3]{<<, >>},
    morekeywords = [4]{++},
    morekeywords = {List, Object},
    basicstyle=\ttfamily\scriptsize,
    commentstyle=\color{blue}\ttfamily,
    morecomment=[f][\lstbg{red!20}]-,
    morecomment=[f][\lstbg{green!20}]+,
    morecomment=[f][\textit]{@@},
    numbers=left,
    xleftmargin=4mm,
    numbersep=7pt,
    numberstyle=\scriptsize,
    frame=none,
    texcl=false,
    moredelim=[il][\textcolor{pgrey}]{\$\$},
    moredelim=[is][\textcolor{pgrey}]{\%\%}{\%\%}
}

\definecolor{grey}{rgb}{0.7,0.7,0.7}

\newcommand{\lstbg}[3][0pt]{{\fboxsep#1\colorbox{#2}{\strut #3}}}
\lstdefinelanguage{diff}{
  basicstyle=\ttfamily\footnotesize,,
  morecomment=[f][\lstbg{red!20}]-,
  morecomment=[f][\lstbg{green!20}]+,
  morecomment=[f][\textit]{@@},
  texcl=false
}

\newtcolorbox{definitionA}[1][]{
  enhanced,
  colback=white,
  colframe=blue!60!black,
  arc=3pt,
  boxrule=0.8pt,
  left=6pt,
  right=6pt,
  top=8pt,
  bottom=8pt,
  fonttitle=\bfseries,
  title=Definition,
  #1
}

\newtcolorbox{definitionC}[1][]{
  enhanced,
  colback=gray!3!white,
  colframe=gray!60!black,
  arc=2pt,
  boxrule=0.5pt,
  left=8pt,
  right=8pt,
  top=10pt,
  bottom=10pt,
  fonttitle=\bfseries,
  coltitle=black,
  title=Definition,
  #1
}

\definecolor{funcgreen}{RGB}{0,100,0}

\lstdefinelanguage{Rust}{
  sensitive=true,
  morekeywords={use,struct,fn,let,mut,unsafe,pub,impl,self,Self,return},
  morecomment=[l]{//},
  morecomment=[s]{/*}{*/},
  morestring=[b]"
}

\def\sys{{\textsc{TRust}}\xspace}
\def\sysb{{\textsc{\textbf{TRust}}}\xspace}

\newcommand{\yp}[1]{\todo[color=green!40]{#1}}

\newcommand{\zhe}[1]{\textcolor{blue}{#1}}
\newcommand{\modify}[1]{\textcolor{blue}{#1}}
\newcommand{\allissuecnt}{308}
\newcommand{\issuecnt}{246}
\newcommand{\faultcnt}{46}
\newcommand{\confirmedfaultcnt}{17}
\newcommand{\fixedfaultcnt}{0}
\newcommand{\fpcnt}{0}
\newcommand{\rootcausecnt}{six}
\newcommand{\symptomcnt}{four}
\newcommand{\fixstrategycnt}{seven}
\newcommand{\findingcnt}{ten}
\newcommand{\lessoncnt}{four}
\newcommand{\clustercnt}{102}
\newcommand{\validtransformation}{15}

\newcommand{\numtool}{six}
\newcommand{\vulcnt}{320}
\newcommand{\statfier}{Statfier\xspace}
\newcommand{\lockbud}{LockBud\xspace}
\newcommand{\semgrep}{Semgrep\xspace}
\newcommand{\rudra}{Rudra\xspace}
\newcommand{\prusti}{Prusti\xspace}
\newcommand{\miri}{MIRI\xspace}
\newcommand{\mirai}{MIRAI\xspace}
\newcommand{\mirchecker}{MirChecker\xspace}

\newcommand{\sonar}{SonarQube\xspace}
\newcommand{\pmd}{PMD\xspace}
\newcommand{\sbugs}{SpotBugs\xspace}
\newcommand{\cstyle}{CheckStyle\xspace}
\newcommand{\infer}{Infer\xspace}
\newcommand{\soot}{Soot\xspace}
\newcommand{\eprone}{Error-prone\xspace}

\definecolor{rustKeyword}{HTML}{008000}
\definecolor{rustFunction}{HTML}{0066B3}
\definecolor{rustType}{HTML}{B00060}
\definecolor{rustBuiltin}{HTML}{3333AA}
\definecolor{rustComment}{HTML}{888888}

\lstdefinelanguage{RustPaper}{
  sensitive=true,
  morekeywords={
    use,struct,fn,let,mut,unsafe,pub,impl,
    enum,match,if,else,return,for,while,
    loop,const,static,mod,trait,where,
    as,ref,move,self,Self,in,break,continue
  },
  morekeywords=[2]{
    str,char,bool,
    u8,u16,u32,u64,u128,usize,
    i8,i16,i32,i64,i128,isize,
    f32,f64
  },
  morekeywords=[3]{PwEntry},
  morecomment=[l]{//},
  morecomment=[s]{/*}{*/},
  morestring=[b]"
}

\lstdefinestyle{rustpaper}{
  language=RustPaper,
  basicstyle=\normalfont\ttfamily\footnotesize,
  keywordstyle=\color{rustKeyword}\bfseries,
  keywordstyle=[2]\color{rustBuiltin}\bfseries,
  keywordstyle=[3]\color{rustType}\bfseries,
  commentstyle=\color{rustComment}\mdseries\upshape,
  stringstyle=\color{black}\mdseries,
  identifierstyle=\color{black},
  emph={make_entry,print_and_free,main,println},
  emphstyle=\color{rustFunction}\bfseries,
  numbers=left,
  numberstyle=\normalfont\rmfamily\scriptsize\color{black},
  numbersep=4pt,
  stepnumber=1,
  firstnumber=1,
  columns=fullflexible,
  keepspaces=true,
  showstringspaces=false,
  tabsize=2,
  breaklines=true,
  breakatwhitespace=false,
  breakautoindent=true,
  breakindent=1em,
  frame=none,
  xleftmargin=0.8em,
  aboveskip=5pt,
  belowskip=3pt
}

\begin{abstract}
Rust has become a popular systems programming language in recent years because of its strong guarantees on memory and concurrency safety.
Many types of system software have been re-implemented in Rust.
However, Rust programs can still suffer from common memory and concurrency vulnerabilities due to unsafe code and misunderstandings of its complex mechanisms.
To mitigate such vulnerabilities, developers often rely on static application security testing (SAST) tools.
However, the effectiveness of existing Rust SAST tools on real-world vulnerabilities remains underexplored.
In this paper, we first conduct an empirical evaluation of mainstream Rust SAST tools using security vulnerabilities identified in real-world Rust applications.
We systematically assess their effectiveness, uncover common weaknesses, and analyze the root causes behind them.
Motivated by the study, we propose \sys{}, a large language model (LLM)-assisted automated testing framework.
\sys{} automatically analyzes the official Rust reference to derive and implement semantics-preserving transformations.
It leverages these transformations to generate mutant programs for evaluating Rust SAST tools and uncovering potential faults within them.
Overall, \sys{} has revealed \faultcnt{} faults across six popular SAST tools, \confirmedfaultcnt{} of which have been confirmed.
In addition, our study yields \findingcnt{} empirical findings and \lessoncnt{} practical lessons that provide actionable guidance for developing more effective, robust, and maintainable Rust SAST tools.
\end{abstract}


\section{Introduction}
\label{sec:intro}
Rust has become a popular systems programming language and is widely used in domains including operating systems~\cite{redox}, compilers~\cite{swc}, and databases~\cite{tikv}.
Rust ensures memory and concurrency safety through novel mechanisms such as ownership and borrow checking.
However, developers may misunderstand or misuse Rust intrinsic language features, especially \texttt{unsafe} for low-level interactions, leaving Rust programs vulnerable to security issues.
SAST tools can detect such vulnerabilities through static analysis without executing the programs, making them well suited to various Rust applications.
However, unlike traditional programming languages such as C/C++ and Java, Rust has its own distinctive features.
These features can affect the essential information required by static analysis; consequently, ignoring them can greatly reduce the accuracy of SAST tools.
Figure~\ref{fig:ownership_example} presents a Rust program example related to the \texttt{ownership} mechanism.
Some SAST tools, such as \lockbud{}~\cite{lockbud2020pldi}, incorrectly treat the underlying allocation of \texttt{dir} as if it were freed by the call to \texttt{into\_raw} at lines 8--9.
In fact, \texttt{into\_raw} only transfers ownership from \texttt{CString} to a raw pointer without freeing the underlying memory.
As a result, SAST tools may report a use-after-free (UAF) warning at lines 15--16 for this example, which is a false positive.

\begin{figure}[tb]
\centering
\begin{minipage}{\linewidth}
\begin{minted}[fontsize=\footnotesize,numbersep=4pt,xleftmargin=0.8em]{rust}
use std::ffi::CString;
use std::os::raw::c_char;
struct PwEntry {
 pw_dir: *mut c_char,
}
fn make_entry(path: &str) -> PwEntry {
 let dir = CString::new(path).unwrap();
 let dir_ptr = dir.into_raw();
 PwEntry { pw_dir: dir_ptr }
}
fn print_and_free(entry: PwEntry) {
 unsafe {
  // from_raw() takes ownership back into a 
  // CString, which will then be freed on drop
  let dir = CString::from_raw(entry.pw_dir);
  println!("dir = {:?}", dir);
  // dir is dropped here; the underlying 
  // allocation is freed exactly once
 }
}
fn main() {
 let entry = make_entry("/home/user");
 print_and_free(entry);
}
\end{minted}
\end{minipage}
\caption{A False Positive in UAF Vulnerability Detection}
\label{fig:ownership_example}
\end{figure}

This example illustrates a concrete challenge for applying SAST tools to Rust: if a tool does not correctly model Rust-specific semantics such as ownership transfer, it may misinterpret memory states and produce false alarms.
Since Rust provides several language features that differ substantially from those of C/C++ and Java, it is important to understand how existing SAST tools handle such features in practice.
To this end, we conduct the first comprehensive empirical study evaluating the capability, effectiveness, and limitations of existing SAST tools for Rust.
Particularly, our study aims to answer the following research questions (RQs):
\begin{itemize}[leftmargin=*]
\item \textbf{RQ1:} What vulnerability types is each SAST tool expected to detect?
This RQ helps us understand the capability of each SAST tool in terms of vulnerability coverage and reveals whether different tools exhibit complementary strengths in detecting specific categories of security issues.

\item \textbf{RQ2:} How do SAST tools perform on real-world projects?
This RQ assesses the effectiveness of these tools in detecting real-world vulnerabilities, focusing on false positive (\textbf{FP}) and false-negative (\textbf{FN}) rates in realistic software development scenarios.

\item \textbf{RQ3:} What are the underlying causes of inaccurate analysis results?
This RQ investigates the limitations of SAST tools by analyzing FPs and FNs and identifies the key factors that affect their effectiveness.

\end{itemize}

To address the research questions, we manually collected \vulcnt{} vulnerabilities from four popular vulnerability sources and evaluated six popular SAST tools, including \semgrep{}, \lockbud{}, \prusti{}, \rudra{}, \mirai{}, and \mirchecker{}, on this dataset.
Through this investigation, we summarized ten key findings from our analysis and derived four lessons for improving Rust SAST tools.
Based on these findings, we proposed \sys{}, an LLM-assisted testing framework that can identify and apply semantics-preserving program transformations derived from the Rust Reference~\cite{rust_ref}.
On the six aforementioned SAST tools, \sys{} successfully detected \faultcnt{} previously unknown faults.
We have reported all detected faults to the corresponding developers, and \confirmedfaultcnt{} of them have been confirmed to date,
which clearly demonstrates the value of \sys{} and the findings summarized in this study.

Prior studies have shown that semantics-preserving transformations can effectively expose faults in a wide range of SAST tools~\cite{statfier, tdsc2024query, glfuzz}.
However, these approaches rely on predefined transformation patterns, limiting their ability to adapt to evolving language features. 
Moreover, existing evaluations of SAST tools have predominantly focused on established languages such as C/C++~\cite{tdsc2024query}, Java~\cite{fse2023sast, issre2004compare}, and Solidity~\cite{issta2020smartcontract, oakland2024smartcontract_sast, fse2024sast},
leaving newer programming languages insufficiently examined.
This gap is particularly significant in the case of Rust, a modern systems programming language with unique features such as ownership and borrowing that fundamentally shape program behavior. 
Although Rust is designed to enforce strong memory-safety guarantees, vulnerabilities can still emerge from unsafe code, developer misconceptions, and complex language interactions. As a result, understanding the effectiveness and limitations of Rust-specific SAST tools is both necessary and timely, yet remains largely underexplored.

Overall, this paper makes the following contributions:
\begin{itemize}[leftmargin=*]
\item To the best of our knowledge, we present the first empirical study of Rust SAST tools.
Our study is based on \vulcnt{} real-world vulnerabilities collected from four popular security advisory databases.
Using these samples, we analyze the effectiveness of these tools and derive \findingcnt{} key findings and \lessoncnt{} lessons for future SAST tool design.

\item Motivated by these findings, we propose \sys{}, a novel testing framework that synthesizes semantics-preserving program transformations based on the language reference and automatically detects faults in SAST tools.

\item We evaluate \sys{} on six SAST tools and uncover \faultcnt{} previously unknown faults, \confirmedfaultcnt{} of which have been confirmed by the corresponding developers. The experimental data and source code of \sys{} are publicly available at \href{https://anonymous.4open.science/r/TRust-TSE-Artifact-9A58}{\textcolor{blue}{Link}}.

\end{itemize}

\section{Background: The Rust Programming Language}
\label{sec:background}
Rust is a modern systems programming language designed to provide memory safety, thread safety, and high performance without relying on garbage collection.
One of its key feature is an ownership-based type system that enforces ownership, borrowing, and lifetime rules at compile time, preventing common errors such as dangling pointers and data races.
Meanwhile, Rust supports zero-cost abstractions,
allowing high-level expressiveness with performance comparable to that of C and C++.
These properties have driven its adoption in security-critical and performance-sensitive domains.

\noindent \textbf{Expressive Syntax.}
Rust includes several distinctive syntactic features, such as pattern matching with \texttt{match}, destructuring bindings, and algebraic data types defined by \texttt{enum}.
These constructs enable concise and expressive handling of complex control flow and data structures.
In addition, trait-based generics and macros provide flexible abstraction mechanisms while preserving runtime efficiency.

\noindent \textbf{Ownership and Borrowing.}
Rust manages memory through its ownership model, in which each value has a unique owner and its lifetime is tied to the scope of owner.
When the owner goes out of scope, the value is automatically dropped, enabling safe and efficient resource management without a garbage collector.
To allow access without transferring ownership, Rust further supports borrowing through references.
Borrowing permits either shared immutable access or exclusive mutable access under compile-time checks, helping prevent issues such as use-after-free and dangling pointers.



\noindent \textbf{Concurrency Safety.}
Rust enforces thread safety through its ownership system and the \texttt{Send}/\texttt{Sync} traits, which determine whether data can be transferred between threads or shared across them.
Traits in Rust are similar to interfaces in Java, in that they define shared behavior that types can implement, often with optional default methods.
Together with borrowing rules, these mechanisms prevent data races at compile time.
\section{An Empirical Study on Rust SAST Tools}
\label{sec:study}

\subsection{Selection Criteria for Target SAST Tools}
\label{sec:toolselection}
This section outlines the criteria used to select target SAST tools. The selection process is guided by three key dimensions:
(1) \textbf{Diversity:} We consider SAST tools from both industry and academia to ensure representation of different development backgrounds;
(2) \textbf{Popularity:} Preference was given to open-source tools with a minimum of 100 stars on their code repository, indicating a certain level of community adoption and trust;
(3) \textbf{Capability:} The selected tools are primarily designed for security vulnerability detection, aligning with the focus of this study.
Based on these criteria, we choose \semgrep{}~\cite{semgrep}, \lockbud{}~\cite{lockbud2020pldi}, \prusti{}~\cite{prusti}, \rudra{}~\cite{rudra}, \mirai{}~\cite{mirai}, and \mirchecker{}~\cite{mirchecker} as our research targets.
\semgrep{}~\cite{semgrep} is a fast SAST tool that searches code, finds bugs, and enforces secure guardrails and coding standards.
\lockbud{}~\cite{lockbud2020pldi} is a SAST tool designed to detect various concurrency and memory safety issues in Rust code, including double-locks, conflicting lock orders, atomicity violations, and use-after-free.
\prusti{}~\cite{prusti} is a static verifier for Rust, based on the Viper verification infrastructure.
It can verify absence of integer overflows and panics, proving that statements e.g. \texttt{panic!()} are unreachable.
\rudra{}~\cite{rudra} is a SAST tool to detect common undefined behaviors in Rust programs. It is capable of analyzing single Rust packages as well as all the packages on \texttt{crates.io}.
\mirai{}~\cite{mirai} is a SAST tool for Rust that functions both as a linter for unintended panics and as a verifier for correctness, API documentation, and security properties such as taint analysis and constant-time execution.
\mirchecker{}~\cite{mirchecker} can analyze the MIR generated by the Rust compiler and emit diagnostic messages.
It is based on the abstract interpretation.
\semgrep{} and \mirai{} are industrial SAST tools, while the other four originate from academia.

\subsection{Benchmark Construction}
\label{sec:benchmark}
\begin{figure}[tb]
\centering
\includegraphics[width=0.9\linewidth]{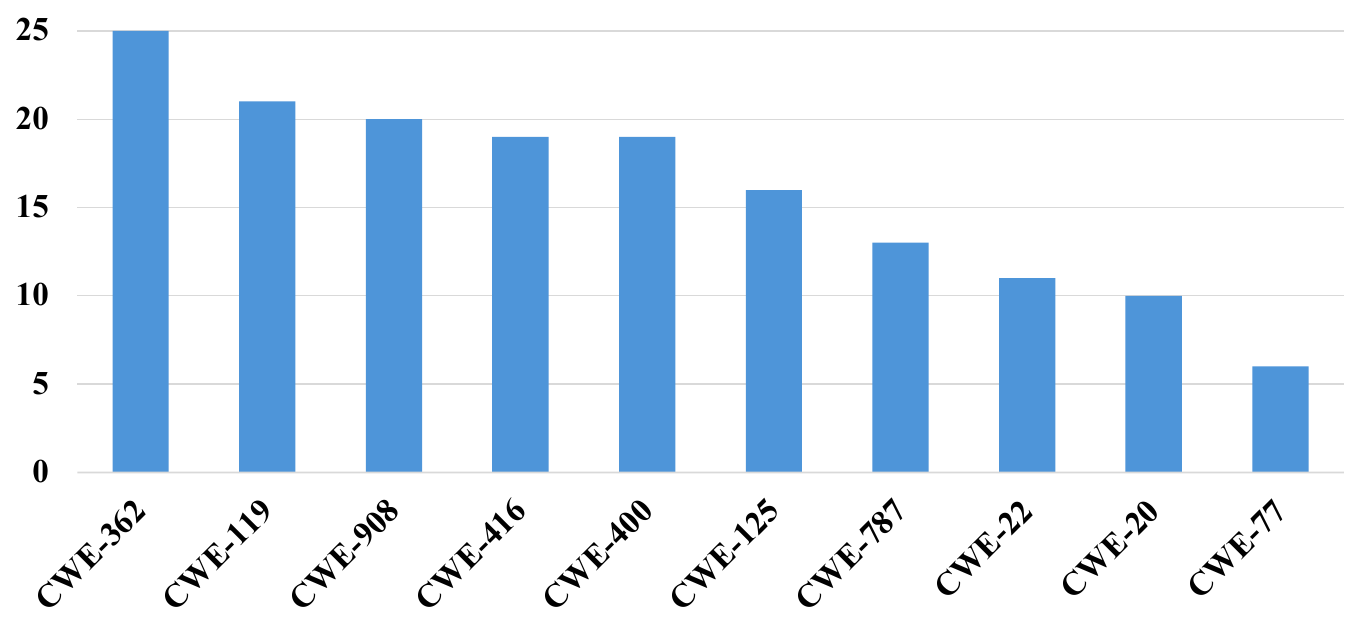}
\caption{Distribution of Vulnerability Types in Our Constructed Benchmark}
\label{fig:cwe-distribution}
\end{figure}

To evaluate the effectiveness and limitation of Rust SAST tools, we first construct a benchmark comprising a diverse set of security vulnerabilities in Rust programs.
Publicly disclosed Rust vulnerabilities are distributed across multiple advisory and vulnerability databases.
Since these sources differ in scope and coverage, relying on a single database may lead to incomplete vulnerability collection.
Following prior literature~\cite{benchmark}, we examine four widely used vulnerability data sources: (1) the RustSec Advisory Database (RustSec)~\cite{rustsec}, (2) GitHub Security Advisories (GSA)~\cite{gsa}, (3) the National Vulnerability Database (NVD)~\cite{nvd}, and (4) the Global Security Database (GSD)~\cite{gsd}.
RustSec~\cite{rustsec} is a repository of security advisories for Rust crates published on \texttt{crates.io} and maintained by the Rust Secure Code Working Group. GSA~\cite{gsa} collects security advisories for open-source software, including both CVEs and GitHub-originated advisories. NVD~\cite{nvd} is maintained by the National Institute of Standards and Technology (NIST) and provides standardized vulnerability information for a broad range of software and hardware systems. GSD~\cite{gsd} aggregates vulnerability and advisory information from multiple public sources.
We collect vulnerability samples from these four data sources over the past five years.
We then retain only samples that include both a CWE ID and the commit ID corresponding to the vulnerable version.
In total, we obtain \vulcnt{} vulnerability samples, covering 84 vulnerability types identified by CWE IDs.
Figure~\ref{fig:cwe-distribution} shows the distribution of the top 10 most common vulnerability types.
CWE-362 (Race condition) is the most frequent category, with 25 samples, followed by CWE-119 (Buffer overflow) and CWE-908 (Use of uninitialized resource) with 21 and 20 samples, respectively.
Overall, the benchmark covers a diverse set of vulnerability types, although their distribution is somewhat imbalanced across categories.

\subsection{RQ1: Analysis Capability of Rust SAST Tools}
\label{sec:rq1}
\begin{figure}[t]
\centering
\includegraphics[width=0.7\linewidth]{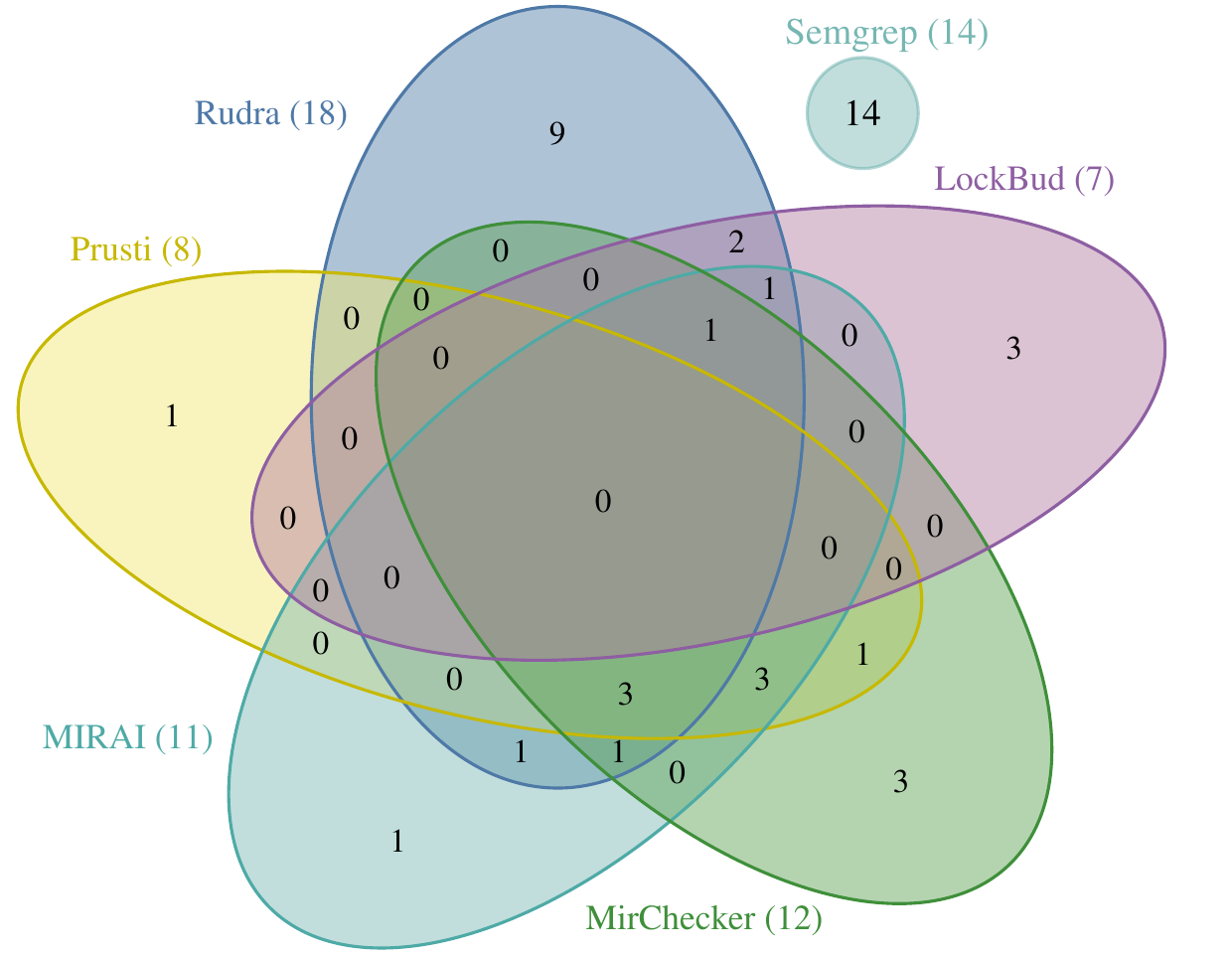}
\caption{Cross-Tool Overlap of Detected CWE IDs}
\label{fig:rq1-distribution}
\end{figure}

In RQ1, we conduct an in-depth analysis of the types of CWE-IDs supported by each SAST tool.
Although different SAST tools may target the same type of vulnerability, their vulnerability detectors are often named differently.
Therefore, mapping these detectors to their corresponding CWE-IDs not only helps us uniformly understand the vulnerability analysis capabilities of different SAST tools, but also enables us to select appropriate vulnerability samples in the following research questions within the scope of each tool's capabilities.
For each SAST tool, we first exclude rules unrelated to security. We then examine whether the documentation specifies the CWE-IDs associated with these rules.
If such information is unavailable, we manually map the rules to CWE-IDs based on their descriptions.
Overall, Table~\ref{tab:rq1-distribution} summarizes the number of CWE-IDs supported by each analyzed SAST tool.
Among these tools, \rudra{} covers the largest number of vulnerability types, largely due to its unique detection logic.
Figure~\ref{fig:rq1-distribution} shows a Venn diagram that illustrates the overlap in detection capabilities among different tools.
As shown in the figure, \semgrep{} supports the second-highest number of vulnerability types, it covers CWE-ID types that are substantially different from those covered by other tools, without overlap.
Other SAST tools mainly focus on the memory and concurrency related vulnerabilities, such as CWE-415 (double free) and CWE-667 (improper locking).

\begin{table}[tb]
\caption{Number of Supported CWE-IDs in each SAST Tool}
\centering
\footnotesize
\setlength{\tabcolsep}{2pt}
\begin{tabular}{l|r|r|r|r|r|r}
\toprule
Tool & Rudra & LockBud & MirChecker & MIRAI & Semgrep & Prusti \\
\hline
\# CWE-ID & 18 & 7 & 12 & 9 & 14 & 8 \\
\bottomrule
\end{tabular}
\label{tab:rq1-distribution}
\end{table}

\subsection{RQ2: Effectiveness of Rust SAST Tools}
\label{sec:rq2}
In this RQ, we aim to evaluate the effectiveness of each SAST tool to understand their performance and usability.
Our approach is to run the SAST tools on the benchmark constructed in Section~\ref{sec:benchmark}.
Then, we evaluate effectiveness from two perspectives: false positives (FPs) and false negatives (FNs).
Following prior manual-analysis studies~\cite{annatester, shen2021dlbugstudy}, we adopted an iterative labeling process to ensure the reliability of the analysis results.
First, the two authors jointly discussed the labeling guidelines and annotated 5\% of the cases as a pilot set.
For FP analysis, they examined each reported warning together with its code context, warning type, description, and, when available, the reported root cause, and labeled it as either a true positive (TP) or an FP.
For FN analysis, they examined the known benchmark vulnerabilities that were not reported by the tools and determined whether each missed case should be labeled as a false negative.
The Cohen's Kappa coefficient~\cite{cohen} for this round was 0.67.
After a training discussion to refine the criteria, they labeled 10\% of the cases (including the previous 5\%), and the Cohen's Kappa coefficient increased to 0.91.
The remaining cases were then labeled in multiple iterations, during which the Cohen's Kappa coefficient consistently remained above 0.9.
In each iteration, the two authors discussed the cases with inconsistent labels until a consensus was reached.
Finally, all FP and FN cases were labeled consistently.

\subsubsection{FP Analysis}
As a high FP rate is one of the biggest concerns for SAST tool users~\cite{icse2022fprate}, performing FP analysis helps us understand the usability of each tool.
As the constructed benchmark~\ref{sec:benchmark} contains too many repositories for manual FP inspection, we select the top five repositories by star count from each year between 2021 and 2025 for FP analysis, resulting in a total of 25 repositories for evaluation.



\begin{table}[t]
\caption{False Positive Rates of the Evaluated SAST Tools}
\centering
\footnotesize
\setlength{\tabcolsep}{2pt}
\begin{tabular}{l|r|r|r|r|r|r}
\toprule
\textbf{Tool} & \textbf{Rudra} & \textbf{MirChecker} & \textbf{LockBud} & \textbf{Semgrep} & \textbf{MIRAI} & \textbf{Prusti} \\
\hline
\#Warning    & 104 & 392    & 173 & 256 & 456 & 20 \\
\#TP & 12  & 70 & 32  & 108 & 71  & 17 \\
\#FP & 92  & 322    & 141 & 148 & 385 & 3  \\
FP Rate & 88.5\% & 82.1\% & 81.5\% & 57.8\% & 84.4\% & 15\% \\
\bottomrule
\end{tabular}
\label{tab:rq2-fprate}
\end{table}

\begin{table}[tb]
\caption{Top 10 CWE IDs with the highest false positive rates}
\centering
\begin{tabular}{l|l|r|r}
\toprule
\textbf{CWE-ID} & \textbf{Vulnerability Type} & \textbf{FP/Total} & \textbf{FP Rate} \\
\hline
CWE-833 & Deadlock & 15/15  & 100\% \\
CWE-369 & Divide by Zero   & 8/8 & 100\% \\
CWE-787 & Out-of-bounds Write & 4/4 & 100\% \\
CWE-122 & Buffer Overflow  & 4/4 & 100\% \\
CWE-416 & Use After Free   & 38/39  & 97.4\%   \\
CWE-125 & Out-of-bounds Read  & 105/109   & 96.3\%   \\
CWE-190 & Integer Overflow & 248/261   & 95\%  \\
CWE-191 & Integer Underflow   & 31/34  & 91.2\%   \\
CWE-367 & Race Condition   & 113/124   & 91.1\%   \\
CWE-662 & Improper Synchronization & 53/60  & 88.3\% \\
\bottomrule
\end{tabular}
\label{tab:rq2-fp-distribution}
\end{table}

Table~\ref{tab:rq2-fprate} shows the FP rates of all SAST tools.
Among them, \semgrep{} appears to provide a better performance considering the number of detected vulnerabilities and the FP rate.
Although \prusti{} achieves the lowest FP rate, at 15\%, it reports only 20 warnings in total, which is substantially fewer than the other tools.
Excluding these two tools, all remaining tools exhibit very high FP rates (above 80\%), indicating limited usability in real-world settings.
We also investigate the top ten CWE IDs with the highest FP rates across all tools.
Table~\ref{tab:rq2-fp-distribution} illustrates their distribution.
Across all SAST tools, four types of vulnerabilities (Deadlock, Divide by Zero, Out-of-bounds Write, and Buffer Overflow) have no TPs.


\begin{findingbox}
The vulnerability types with the highest FP rates are mostly memory- and concurrency-related.
This suggests that, although these categories are central to Rust security analysis, precisely distinguishing real bugs from safe Rust idioms in these categories remains challenging for current SAST tools.
\end{findingbox}

To better understand the differences between Rust SAST tools and those for traditional programming languages, we compare their FP rates against those reported for six representative C/C++ SAST tools in prior work~\cite{tdsc2024query}.
The reported FP rates are 23.3\% for CodeQL, 54.9\% for Joern, 68.4\% for \semgrep{}, 76.8\% for \sonar{}, 39.1\% for Coverity, and 60.9\% for Clang, corresponding to an average FP rate of 53.9\%.
By comparison, the average FP rate of Rust SAST tools in our study is 80.3\%.
Although these results are drawn from different studies, they still provide useful evidence that Rust SAST tools currently tend to produce more false alarms than those for traditional programming languages.


\begin{findingbox}
Compared with FP rates reported for mature C/C++ SAST tools in prior work, Rust SAST tools in our study show higher FP rates.
While this cross-study comparison should be interpreted cautiously, it suggests that Rust-specific abstractions introduce additional precision challenges for static analysis.
\end{findingbox}

\subsubsection{FN Analysis}
Conducting FN analysis helps us understand the reliability of each tool, as FNs indicate real vulnerabilities that the tool fails to detect, undermining the security assurance it can provide.
To measure the FN rate, we use all collected 320 vulnerabilities as the evaluation set.
We run all SAST tools on these repositories and manually examine how many vulnerabilities they successfully detect based on the labels provided by the vulnerability vendors.

\begin{table}[tb]
\caption{False Negative Rates of the Evaluated SAST Tools}
\centering
\footnotesize
\setlength{\tabcolsep}{2pt}
\begin{tabular}{l|r|r|r|r|r|r}
\toprule
\textbf{Tool} & \textbf{Rudra} & \textbf{MirChecker} & \textbf{LockBud} & \textbf{Semgrep} & \textbf{MIRAI} & \textbf{Prusti} \\
\hline
\# Expected & 320 & 320 & 320 & 320 & 320 & 320 \\
\#FN & 301 & 319 & 319 & 316 & 319 & 319 \\
FN Rate & 94.1\% & 99.7\% & 98.8\% & \% & 97.7\% & 97.7\% \\
\bottomrule
\end{tabular}
\label{tab:rq2-fnrate}
\end{table}


Table~\ref{tab:rq2-fnrate} presents the FN rate of each SAST tool.
The second row reports the number of FNs, and the third row reports the FN rate.
All tools exhibit very high FN rates (more than 90\%).
Notably, although some vulnerabilities may be beyond the detection capabilities of current tools, these samples still reflect the distribution of real-world vulnerabilities.

\subsection{RQ3: Root Causes of Analysis Failures}
\label{sec:rq3}
In RQ2, we observe that all evaluated SAST tools suffer from high FP and FN rates.
However, the reasons behind these high FP and FN rates remain unclear.
In this RQ, we further analyze the FPs and FNs identified in our evaluation to investigate their root causes and understand their limitations.

\subsubsection{Root Causes of FPs}
Based on the analysis of FPs in RQ2, we identify three main root causes:
(1) improper implementation of analysis techniques (45.4\%),
(2) incorrect handling of Rust language features (40.3\%),
and (3) overly simplistic pattern design (14.3\%).

\noindent \textbf{Improper Implementation of Analysis Techniques (495).}
SAST tools typically rely on various static analysis techniques, and we find that issues in three of these techniques account for the majority of cases under this category.
The first is inaccurate pointer analysis (31).
Figure~\ref{fig:pointer_analysis_example} presents an illustrative example of this root cause: in the next iteration, \lockbud{} mistakenly assumes that the \texttt{drop} at line 5 in the previous iteration has released the temporary variable created at line 4.
\begin{figure}[tb]
\centering
\begin{minipage}{\linewidth}
\begin{minted}[fontsize=\footnotesize,numbersep=4pt,xleftmargin=0.8em]{rust}
for (idx, page) in self.pages.iter().enumerate().skip(1) {
  let mut slots = page.slots.try_lock().unwrap();
  let vec = mem::take(&mut slots.slots);
  assert!(vec.as_ptr() == ptr::null());
  drop(vec);
}
\end{minted}
\end{minipage}
\caption{A False Positive Due to Imprecise Pointer Analysis}
\label{fig:pointer_analysis_example}
\end{figure}
The second type is related to incorrect dependency analysis (118), and Figure~\ref{fig:incorrect_dependency_example} shows such an example that \rudra{} reports a warning at line 6 on \texttt{ptr\_i.write(value\_i)} because \texttt{ptr::write} is an unsafe operation whose correctness depends on the destination pointer being valid and properly aligned.
In this case, however, the warning is an FP.
Here, \texttt{ptr\_i} is derived from the storage of \texttt{MaybeUninit<[T; N]>}, cast to \texttt{*mut T}, and advanced element-wise within the array;
\texttt{value\_i} is a value of type \texttt{T} returned by \texttt{initializer(i)}.
Thus, the write targets a properly aligned element slot in the array and does not violate the precondition of \texttt{ptr::write}.
This FP stems from the imprecise MIR-based dependency analysis of \rudra{}.
It conservatively infers an unsafe flow from the generic callback \texttt{initializer(i)} to the sink \texttt{ptr\_i.write(value\_i)} based on control-flow adjacency in the MIR graph, rather than on an actual value dependency that could make the pointer unsafe.
Since the call at line 5 is a predecessor of the \texttt{write} at line 6 in MIR, the analysis propagates taint to the \texttt{write} and emits a warning, even though there is no real dataflow from \texttt{initializer} to \texttt{ptr\_i} that could invalidate the pointer or make the write unsafe.
\begin{figure}[tb]
\centering
\begin{minipage}{\linewidth}
\begin{minted}[fontsize=\footnotesize,numbersep=4pt,xleftmargin=0.8em]{rust}
let mut arraya: core::mem::MaybeUninit<[T; N]> = core::mem::MaybeUninit::uninit();
let mut ptr_i = array.as_mut_ptr() as *mut T;
unsafe {
  for i in 0..N {
    let value_i = initializer(i);
    ptr_i.write(value_i);
    ptr_i = ptr_i.add(1);
  }
  array.assume_init()
}
\end{minted}
\end{minipage}
\caption{A False Positive Caused by Incorrect Dependency Analysis}
\label{fig:incorrect_dependency_example}
\end{figure}
The third type concerns imprecise interval analysis (350). Figure~\ref{fig:incorrect_interval_analysis_example} illustrates this issue: \mirai{} reports a false integer-overflow warning at line 3 for the statement \texttt{self.pos+=1}.
The root cause is the SAST tool treats \texttt{pos} as an unconstrained integer and fails to recognize its semantic role as an iterator index.
In this example, \texttt{pos} is used to access elements via \texttt{.get(self.pos)}, which implicitly bounds its value by the size of the underlying container.
As a result, an actual overflow would require an unrealistically large container (e.g., on the order of 64GB), rendering the warning spurious.
To reduce FPs in interval analysis, developers should consider the implicit value ranges of specific APIs and variables, e.g., vector-length APIs.

\begin{figure}[t]
\centering
\begin{minipage}{\linewidth}
\begin{minted}[fontsize=\footnotesize,numbersep=4pt,xleftmargin=0.8em]{rust}
fn next(&mut self) -> Option<Self::Item> {
  let ret = self.inner.inner.get(self.pos)
            .map(Event::from_sys_event_ref);
  self.pos += 1;
  ret
}
\end{minted}
\end{minipage}
\caption{A False Positive Due to Imprecise Interval Analysis}
\label{fig:incorrect_interval_analysis_example}
\end{figure}

\begin{findingbox}
Many FPs are rooted in precision limitations of core analysis components rather than only in overly broad detection rules.
In particular, interval analysis, dependency analysis, and pointer analysis are the main contributors, indicating that rule refinement alone may be insufficient without more precise semantic modeling.
\end{findingbox}



\noindent \textbf{Incorrect Handling of Rust Language Feature (440).}
Specifically, we divide this category into three subcategories: concurrency mechanism, ownership semantics, and Rust-specific syntactic construct (see Section~\ref{sec:background} for more details).
Among these, issues related to concurrency mechanisms are the most prevalent, accounting for 208 cases.
We further identify three root causes within this subcategory.
The first root cause is the neglect of lock-related operations (93/208).
For example, in the vulnerability CVE-2024-41815, \lockbud{} incorrectly reports a double-lock vulnerability because it fails to consider the \texttt{drop} operation that occurs between the first and second lock acquisitions.
The second-most issues are related to unawareness of lock structure (66/208).
Figure~\ref{fig:lock_struct} shows such an example that \rudra{} reports a warning on \texttt{UnparkMutex} because the type is unsafely implemented as \texttt{Sync} under the bound \texttt{D: Send}, without requiring \texttt{D: Sync}.
This leads \rudra{} to conservatively regard the \texttt{UnsafeCell<Option<D>>} field as potentially unsafe for shared access across threads.
However, this warning can be an FP, since \texttt{UnparkMutex} is a lock-like synchronization primitive whose internal locking protocol is intended to guarantee the thread safety of its protected data.
\begin{figure}[htbp]
\centering
\begin{minipage}{\linewidth}
\begin{minted}[fontsize=\footnotesize,numbersep=4pt,xleftmargin=0.8em]{rust}
unsafe impl<D: Send> Send for UnparkMutex<D> {}
unsafe impl<D: Send> Sync for UnparkMutex<D> {}
\end{minted}
\end{minipage}
\caption{A False Positive Caused by Failure to Recognize the Lock Structure}
\label{fig:lock_struct}
\end{figure}
The third type is related to the ignorance of the control of entering the critical section using THREAD\_ID (49/208).
Figure~\ref{fig:thread_id} shows such an example that \lockbud{} reports a potential concurrency issue in lines 2--5, but this code does not actually exhibit a multi-threading bug.
Specifically, the value loaded at line 3 is compared with the ID of the current thread, and the \texttt{store} operation is executed only when the two are equal.
Since each thread has a unique thread ID, at most one thread can satisfy this condition and enter the \texttt{if-true} branch.
Therefore, concurrent execution of the block by multiple threads is impossible, and no real concurrency issue exists here.
To reduce such FPs, SAST developers should take Rust-specific locking patterns into account, such as lock-related operations and structures.

\begin{figure}[htbp]
\centering
\begin{minipage}{\linewidth}
\begin{minted}[fontsize=\footnotesize,numbersep=4pt,xleftmargin=0.8em]{rust}
let caller = THREAD_ID.with(|id| *id);
let owner = self.owner.load(Ordering::Acquire);
if caller == owner {
  self.owner.store(THREAD_ID_INUSE, Ordering::Release);
  return self.guard_owned(caller);
}
\end{minted}
\end{minipage}
\caption{An FP of Concurrency Bug due to THREAD\_ID}
\label{fig:thread_id}
\end{figure}

\begin{findingbox}
Rust-specific concurrency mechanisms are a major source of FPs.
Our results suggest that current tools can benefit from more precise modeling of lock lifetimes, lock-like abstractions, and THREAD\_ID-based access-control patterns, which are commonly used to encode synchronization semantics in Rust programs.
\end{findingbox}


The second category of Rust features involves the ownership semantics (160).
SAST tools may misinterpret the ownership status of certain objects and consequently report false warnings. Figure~\ref{fig:ownership_fp} presents such an example: \lockbud{} incorrectly assumes that the object is released by \texttt{into\_raw()} at line 2 and therefore reports a use-after-free (UAF) warning at line 4, as it considers the use of \texttt{pw\_dir} to occur after the release.
In fact, \texttt{into\_raw()} transfers ownership rather than releasing the object, which leads to an FP UAF warning in \lockbud{}.
The third root cause is the improper handling of Rust syntactic constructs (72), such as \texttt{match} expressions and \texttt{macros}.
These constructs may obscure control flow or synchronization behavior, causing SAST tools to analyze them imprecisely and report false warnings, e.g., when a program acquires a lock and handles the result through a \texttt{match} expression, a tool may incorrectly infer that the lock is not properly released along some execution path, even though the Rust RAII mechanism guarantees safe cleanup.

\noindent \textbf{Overly simplistic pattern design (156).}
The final root cause is that many patterns rely on only a single API.
Take \semgrep{} as an example: it often detects path traversal vulnerabilities based solely on read-related APIs such as \texttt{std::fs::read}.
However, these APIs are benign in most cases, and \semgrep{} does not leverage sufficient contextual information to determine whether a reported issue is a real vulnerability, resulting in many FPs.
Although these rules can be manually disabled, developers cannot simply discard them if they still need to detect this type of vulnerability.

\begin{figure}[tb]
\centering
\begin{minipage}{\linewidth}
\begin{minted}[fontsize=\footnotesize,numbersep=4pt,xleftmargin=0.8em]{rust}
let dir = match u.dir.into_os_string().into_string() {
Ok(s)=>CString::new(s.as_str()).unwrap().into_raw(), ... };
Self { pw_name: name, pw_dir: dir, ... }
\end{minted}
\end{minipage}
\caption{A \lockbud{} FP of UAF caused by the Misunderstanding of Ownership}
\label{fig:ownership_fp}
\end{figure}

\subsubsection{Root Causes of FNs}
In this section, we investigate the root causes of FNs, and conclude three main root causes: (1) misalignment between detection rules and real-world vulnerabilities (42.2\%), (2) unexpected analysis failure (41.8\%), (3) lack of support for analyzing Rust language features (15.9\%).

\noindent \textbf{Misalignment Between Detection Rules and Real-World Vulnerabilities (810).}
The most common root cause of FNs is the CWE IDs covered by current SAST tools only partially overlap with the security vulnerabilities found in real-world repositories.
This limited overlap indicates a mismatch between the predefined detection scope of tools and the vulnerability patterns that actually occur in practice.
Consequently, many real-world vulnerabilities fall outside the detection scope and remain undetected, thereby contributing to FNs.
For each SAST tool, Table~\ref{tab:rq3-analysis-scope} reports the proportion of repositories containing vulnerabilities outside its detection scope.
The \textit{OOS} (out-of-scope) row indicates the number of repositories whose vulnerabilities are not covered by the corresponding SAST tool.
The average out-of-scope rate reaches 42.2\% (811/1920), indicating that a substantial proportion of all evaluated repositories contain vulnerabilities beyond the detection scope of the evaluated SAST tools.
In addition, all evaluated SAST tools support 42 unique CWE-based vulnerability types (shown in Table~\ref{tab:rq1-distribution}) and can analyze all repositories without reliability issues.
However, their coverage of real-world vulnerabilities remains limited: among the repositories we studied, only 31 contain vulnerability types that are supported by the evaluated SAST tools, whereas the remaining 53 fall outside the tools’ coverage scope.

\begin{findingbox}
A substantial portion of FNs comes from vulnerabilities that fall outside the detection scope of current tools.
This indicates that improving Rust SAST effectiveness requires not only more precise analyses, but also checker portfolios that better reflect the vulnerability types observed in real-world Rust projects.
\end{findingbox}


\begin{table}[tb]
\caption{Out-of-Scope (OOS) Repos and Rates for SAST Tools}
\centering
\footnotesize
\setlength{\tabcolsep}{2pt}
\begin{tabular}{l|r|r|r|r|r|r}
\toprule
\textbf{Tool} & \textbf{Rudra} & \textbf{MirChecker} & \textbf{LockBud} & \textbf{Semgrep} & \textbf{MIRAI} & \textbf{Prusti} \\
\hline
OOS & 154 & 16 & 199 & 309 & 115 & 18\\
Total Repo & 320 & 320 & 320 & 320 & 320 & 320\\
OOS Rate & 48.1\% & 5\% & 62.2\% & 96.6\% & 35.9\% & 5.6\%\\
\bottomrule
\end{tabular}
\label{tab:rq3-analysis-scope}
\end{table}

\noindent \textbf{Unexpected Analysis Failure (803).}
The second root cause of FN is the unexpected failure during SAST analysis, including crash and hang.
Table~\ref{tab:rq3-analsis-failure} shows the number of issues caused by this root cause.
According to our investigation, one major reason is \emph{compiler front-end failure}.
Some SAST tools depend on nightly Rust compilers or unstable compiler internals, which may break when analyzing repositories developed for different compiler versions.
For example, \prusti{} fails when its underlying compiler encounters removed Rust attributes, leading to premature termination of the analysis.
Another important reason is \emph{toolchain incompatibility between the SAST tool and the analyzed repository}.
As Rust libraries evolve, certain data structure layouts may change across versions.
For instance, developers changed the size of \texttt{SocketAddrV4} from 128 bits to 48 bits.
When the repository performs a \texttt{transmute} between \texttt{SocketAddrV4} and \texttt{libc::sockaddr\_in}, the mismatch between the source type (48 bits) and the target type (128 bits) causes tools such as \rudra{} and \prusti{} to crash.
In such cases, the analysis cannot be completed, and the corresponding vulnerabilities are missed, resulting in FNs.
A third reason is \emph{incomplete support in the analysis engine for Rust-specific constructs}.
Some tools terminate unexpectedly because they cannot correctly model language constructs such as closures and iterators.
For example, a faulty implementation may always select the first closure parameter during analysis and then crash when processing the second parameter due to the \texttt{Option}-based representation.
Similarly, some tools are unable to analyze Rust code that uses iterators.
As observed in our study, \prusti{} explicitly reports that iterators are not fully supported yet, and thus fails on iterator-based loops.
In these cases, the SAST tool terminates before finishing the analysis, causing the corresponding vulnerabilities to be missed and eventually leading to FNs.

\begin{table}[tb]
\caption{Unexpected Analysis Failures of SAST Tools}
\centering
\footnotesize
\setlength{\tabcolsep}{2pt}
\begin{tabular}{l|r|r|r|r|r|r}
\toprule
\textbf{Tool} & \textbf{Rudra} & \textbf{MirChecker} & \textbf{LockBud} & \textbf{Semgrep} & \textbf{MIRAI} & \textbf{Prusti}\\
\hline
\# Crash & 40 & 301 & 55 & 0 & 123 & 284\\
\# Hang & 0 & 0 & 0 & 0 & 0 & 12\\
\# Success & 280 & 19 & 265 & 320 & 197 & 24\\
Failure Rate & 12.5\% & 94.1\% & 17.2\% & 0\% & 38.4\% & 92.5\%\\
\bottomrule
\end{tabular}
\label{tab:rq3-analsis-failure}
\end{table}

\begin{findingbox}
Unexpected analysis failures account for 41.8\% of FNs, suggesting that analysis robustness is not merely an engineering concern but an important component of SAST effectiveness.
This is particularly relevant for Rust, where tools often depend on compiler internals and need to track a rapidly evolving toolchain.
\end{findingbox}



\noindent \textbf{Lack of support for analyzing Rust language features (307).}
We classify this root cause into three categories: ownership semantics, concurrency mechanism, and syntactic construct.
An insufficient understanding of these features may lead SAST tool developers to implement detection rules incorrectly, resulting in FNs.
Figure~\ref{fig:ownership_fn} illustrates an UAF FN related to Rust’s ownership model and the use of \texttt{and\_then}.
The FN arises because the SAST tool fails to precisely model value propagation and ownership transfer across chained combinators gene.
In particular, it does not connect the \texttt{ptr} received by the second closure with the value returned by the first closure, nor does it track that this pointer aliases memory owned by the local \texttt{vec} in the previous closure.
As a result, the SAST tool fails to recognize that the pointer used later still refers to memory owned by \texttt{vec}, even though \texttt{vec} has already been dropped.
\begin{figure}[htbp]
\centering
\begin{minipage}{\linewidth}
\begin{minted}[fontsize=\footnotesize,numbersep=4pt,xleftmargin=0.8em]{rust}
struct Dropping<T>(*const T);
impl<T> Drop for Dropping<T> { fn drop(&mut self) {...} }
let result = Some(vec![1, 2, 3]).and_then(|vec| {
    let dropping = Dropping(&vec[0] as *const i32);
    let value = unsafe { *dropping.0 };
    Some(dropping)
}).and_then(|ptr| {
    let value = unsafe { *ptr.0 };
    Some(value)
});
\end{minted}
\end{minipage}
\caption{A false negative for an ownership-related UAF bug}
\label{fig:ownership_fn}
\end{figure}
The second category of this root cause is concurrency mechanism.
Some SAST tools fail to properly model Rust’s concurrency mechanism, which leads to FNs in detecting concurrency-related vulnerabilities.
As shown in Figure~\ref{fig:concurrency_fn}, \texttt{ARCache<K, V>} is unsafely implemented as both \texttt{Send} and \texttt{Sync}, while the generic type \texttt{V} is only constrained by \texttt{Clone} and \texttt{Debug}.
However, to preserve Rust’s concurrency guarantees, \texttt{V} should additionally satisfy Send for the Send implementation and \texttt{Sync} for the \texttt{Sync} implementation.
Without these trait bounds, \texttt{ARCache<K, V>} may be incorrectly treated as safe to transfer or share across threads even when \texttt{V} itself is not thread-safe.
This example illustrates that some existing SAST tools do not adequately reason about Rust’s trait-based concurrency semantics, and therefore miss such unsafe implementations, resulting in FNs.

\begin{figure}[htbp]
\centering
\begin{minipage}{\linewidth}
\begin{minted}[fontsize=\footnotesize,numbersep=4pt,xleftmargin=0.8em]{rust}
unsafe impl<K: Hash + Eq + Ord + Clone + Debug, V: Clone + Debug> Send for ARCache<K, V> {}
unsafe impl<K: Hash + Eq + Ord + Clone + Debug, V: Clone + Debug> Sync for ARCache<K, V> {}
\end{minted}
\end{minipage}
\caption{A false negative for concurrency vulnerability}
\label{fig:concurrency_fn}
\end{figure}

\begin{findingbox}
Rust-specific language features are a notable source of FNs. In particular, ownership propagation across higher-order combinators and trait-based concurrency constraints such as \texttt{Send} and \texttt{Sync} are difficult to model precisely, which may cause tools to miss vulnerabilities whose security conditions are encoded in these semantics.
\end{findingbox}


\section{\sys{} Implementation}
\label{sec:implementation}

The findings of RQ2 and RQ3 point to a key challenge in testing Rust SAST tools: analysis quality heavily depends on whether Rust-specific language features are modeled correctly.
These observations suggest that a useful oracle for testing Rust SAST tools should not merely compare outputs across arbitrary program variants; instead, it should test whether a SAST tool remains stable when the same vulnerability-relevant behavior is expressed through different but semantically equivalent Rust formulations.
The key idea is straightforward: if two programs are equivalent with respect to vulnerability-relevant behavior, then a reliable SAST tool should produce consistent analysis results on both of them.

Following this idea, we design a transformation-based oracle for Rust SAST testing.
Formally, let $A$ denote a Rust SAST tool, and let $P$ be a seed program. We apply a transformation $T$ to obtain a mutant $P' = T(P)$.
The transformation is considered valid only if $P' \equiv P$, where $\equiv$ denotes semantic equivalence with respect to vulnerability-relevant behavior.
Let $\mathit{Res}(A, P)$ denote the analysis result of $A$ on $P$, including the reported warning set and their associated categories.
Our oracle requires that
$P' \equiv P \Longrightarrow \mathit{Res}(A, P') \approx \mathit{Res}(A, P),$
where $\approx$ denotes consistency of analysis results under semantics-equivalent transformation.

Our oracle meets three requirements.
First, it should be \textbf{semantics-preserving}, ensuring that mutants retain the vulnerability-relevant behavior of the originals.
Second, it should be \textbf{language-feature-aware}, as many analysis failures stem from poor handling of Rust-specific constructs.
Third, it should be \textbf{adaptable to language evolution}, since Rust evolves rapidly and a fixed set of handcrafted transformations can quickly become outdated.
These requirements motivate our implementation strategy. Although prior work has shown the effectiveness of semantics-preserving transformations for testing SAST tools~\cite{statfier, glfuzz}, existing approaches typically rely on manually designed transformations, which are insufficient for Rust, where many analysis failures are rooted in language-specific semantics and rapidly evolving idioms (Findings 4, 5, and 7). To address this challenge, we use an LLM to analyze the Rust Language Reference~\cite{rust_ref} (a book that defines features in stable Rust), and automatically construct semantics-preserving, language-feature-aware transformations.
To ensure correctness, we further validate each candidate transformation on executable programs by checking input-output equivalence before using it for testing.
In this way, the oracle serves as the bridge from the empirical failure analysis in RQ2 and RQ3 to a systematic methodology for automated testing.

\begin{figure*}[!htbp]
\centering
\includegraphics[width=\textwidth]{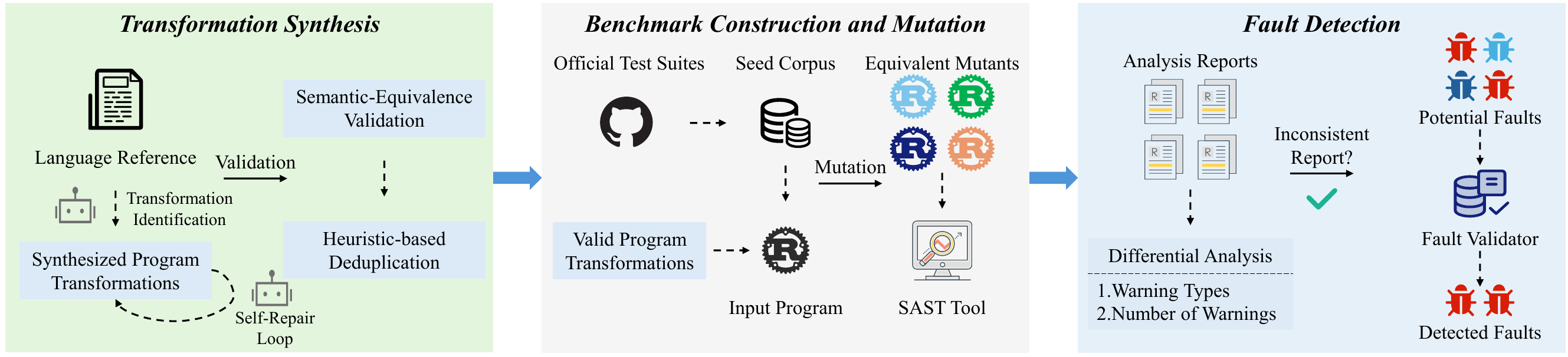}
\caption{Overall Workflow of \sys{}}
\label{fig:workflow}
\end{figure*}

Figure~\ref{fig:workflow} illustrates the overall workflow of \sys{}.
In general, \sys{} consists of three stages.
First, it synthesizes candidate semantics-preserving transformations from the Rust reference~\cite{rust_ref}, and an LLM automatically generates their corresponding implementations.
\sys{} then applies these candidate rules to test programs, checks the semantic equivalence between the resulting programs and the originals, and removes duplicates, yielding a set of validated transformations.
Second, \sys{} applies the retained transformations to seed programs collected from the official test suites and invokes an SAST tool to generate analysis reports for both the seed and mutant programs.
Finally, \sys{} performs differential analysis on these reports and treats inconsistencies as potential faults.


\subsection{LLM-Assisted Synthesis of Program Transformations}
\label{sec:synthesis}
The key challenge in \sys{} is to automatically construct transformations that are both semantics-preserving and aware of Rust language features.
The prompt template used for transformation synthesis is provided in Figure~\ref{fig:prompt_synthesis}.
Specifically, we provide the LLM with the Rust Reference~\cite{rust_ref} as transformation knowledge source, and the LLM is asked to identify language constructs and APIs that admit equivalent formulations, summarize the semantic conditions under which such equivalence holds, and produce transformation rules together with their applicability constraints.
In this way, the specification acts as the semantic boundary of the transformation, while the LLM serves as the mechanism for extracting and instantiating transformation opportunities.
We divide the language reference into 121 page-level segments and feed each segment to the LLM in a separate prompt.

\begin{figure}[!htbp]
\centering
\begin{minipage}{\linewidth}
\begin{userprompt}
\textbf{[Task]:}
Your task is to derive semantics-preserving program transformations from the Rust language reference and generate their corresponding implementations.

\textbf{[Function Instructions]:}
1. Read the provided Rust reference document and understand its content;
2. A transformation is semantics-preserving if observable behaviors remain unchanged for the same input...

\textbf{[Implementation Constraints]:}
1. The method must return \textit{HashMap\textless int[], String\textgreater}.
2. Use byte ranges as keys: $new\ int[]\ \{startByte,\ endByte\}$...

\textbf{[Output Instructions]:}
1. Output exactly one JSON array.
2. The array length must be 1 or 2...

\textbf{[Output Schema]:}
[\{
``transformation\_name'': ``\textless name\textgreater'',
``transformation\_description'': \{
  ``lhs'': ``\textless Rust pattern before transformation\textgreater'',
  ``rhs'': ``\textless Rust pattern after transformation\textgreater''
\},
``example\_code'': ``\textless minimal Rust example\textgreater'',
"java\_searchCandidates": "\textless complete Java method text\textgreater"
\}]

\textbf{[Input]:} Rust Reference Document: \texttt{\{Document\}}
\end{userprompt}
\end{minipage}
\caption{Prompt Template for Synthesizing Program Transformations}
\label{fig:prompt_synthesis}
\end{figure}

Each generated response should contain four parts:
(1) the transformation name;
(2) a description of the generated transformation;
(3) an illustrative code example of the transformation;
and (4) a detailed implementation of the program transformation.
This design differs from traditional program transformation generation approaches in two respects.
First, it targets \emph{equivalent reformulations} rather than arbitrary perturbations, thereby aligning the generated variants with the proposed oracle.
Second, it is \emph{language-feature-aware}: the transformation space is explicitly derived from Rust-specific semantics instead of generic syntactic templates.
As a result, the generated transformations can directly exercise the language features that were shown in RQ3 to be frequent sources of analysis failures, such as ownership movement, iterator abstractions, and expressive language constructs.

\subsection{Validation of Synthesized Program Transformations}
\label{sec:validation}
Automatically generated transformations are valuable only if they are genuinely semantics-preserving.
To eliminate invalid transformations, we introduce a validation mechanism before applying them to seeds.
This validation stage consists of two components.
The first component evaluates candidate transformations on programs with deterministic input-output behavior to determine whether they preserve semantics.
To this end, we use programs from programming contests because they are self-contained, executable, and easy to compare based on observable outputs.
Given a candidate transformation, we instantiate it on such programs to construct seed-mutant program pairs.
We then compile and execute both versions with the same input set and compare their outputs.
A transformation is retained only if the mutant compiles successfully and produces outputs identical to those of the seed for all inputs.
The second component performs heuristic-based deduplication, as the synthesis process may generate duplicate transformations.
Specifically, we first employ an LLM to extract key phrases for each generated transformation and then linearly scan the transformation set.
We use Jaccard similarity to determine whether two transformations belong to the same cluster.
Each resulting cluster is treated as a single program transformation after deduplication.

This validation mechanism serves two purposes.
First, behavioral validation provides an inexpensive dynamic check of whether a candidate transformation changes observable behavior in representative usage scenarios, and it helps refine the applicability conditions of the transformation.
If a candidate transformation fails to preserve behavior on any validation program, we discard it.
Second, heuristic-based deduplication reduces redundant testing efforts.
Because the synthesis process may generate duplicate transformations, applying all of them would increase testing cost.
The deduplication process improves testing efficiency by avoiding repeated validation and redundant execution on transformations that are equivalent.
As the LLM may fail to implement a transformation correctly in a single attempt, we iteratively prompt it to repair potential implementation errors using the template in Figure~\ref{fig:prompt_repair}.
We allow up to ten repair attempts; ultimately,all generated transformation implementations compile without errors.

\begin{figure}[!htbp]
\centering
\begin{minipage}{\linewidth}
\begin{userprompt}
\textbf{[Task]}
Your task is to fix potential implementation errors in semantics-preserving transformations implemented in Java.

\textbf{[Function Instructions]}
1. You will be given:
  - one transformation description section from a markdown file...
2. Revise the Java method based on the transformation description, error logs, and benchmark Rust snippet...

\textbf{[Implementation Constraints]}
1. Keep the method signature exactly as ...
2. Use \textit{edits.put(new int[]\{start, end\},\ replacement)} to record edits...

\textbf{[Output Instructions]}
1. Output only the final Java code, not the full class declaration...

\textbf{[Input]}
Transformation Description: \{\texttt{desc}\},
Current Implementation: \{\texttt{implementation}\},
Error Logs: \{\texttt{logs}\},
Rust Example: \{\texttt{example}\}
\end{userprompt}
\end{minipage}
\caption{Prompt Template for Fixing Synthesized Program Transformations}
\label{fig:prompt_repair}
\end{figure}

\subsection{Mutant Generation and Bug Detection}
\label{sec:mutation}
After validation, the validated transformations are applied to the collected seed programs.
For a seed program $P$, \sys{} first checks whether the corresponding preconditions are satisfied.
If so, the seed program is transformed to produce a mutant $P' = T(P)$.
As a single seed may admit multiple valid transformations, \sys{} can generate multiple mutants from the same seed.
This one-to-many strategy can allow us to examine whether a SAST tool behaves consistently across several semantically equivalent formulations of the same underlying behavior.
As transformations are derived from the language reference, the resulting mutants naturally cover a broader range of Rust idioms than manually written templates.

For each seed/mutant pair $(P, P')$, we run the target SAST tool on both programs and compare the normalized analysis reports.
Let $\mathit{Res}(A, P)$ denote the analysis report of the tool $A$ on program $P$.
Following the previous oracle design, we expect
$P' \equiv P \Longrightarrow \mathit{Res}(A, P') \approx \mathit{Res}(A, P).$
If this consistency condition is violated, we report a potential fault.
In our implementation, result comparison is performed after normalization.
This step abstracts away presentation-level differences such as shifted line numbers or formatting changes introduced by transformations, and focuses on analysis decisions that should remain stable under semantics-preserving transformation.
Typical inconsistency patterns include:
(1) a warning reported on $P$ but not on $P'$,
(2) a warning absent on $P$ but newly introduced on $P'$,
and (3) a change in the vulnerability category or checker decision between the two versions.
Such discrepancies suggest that the SAST tool is affected by non-semantic changes and may mishandle the corresponding language features.




\section{Evaluation}
\label{sec:evaluation}

In this section, we evaluate the effectiveness of \sys{} by answering the following research questions:
\begin{itemize}[leftmargin=*]

\item \textbf{RQ4: How effective is \sys{} in detecting faults in Rust SAST tools?}
This RQ evaluates the practical fault-finding capability of \sys{}. We measure how many unique faults \sys{} exposes in the six target Rust SAST tools, and summarize the types of these faults, including FPs, FNs, and crashes.

\item \textbf{RQ5: How many valid semantics-preserving transformations can \sys{} generate, and how useful are they for fault detection?}
This RQ evaluates our transformation synthesis approach by measuring the number of LLM-generated candidate transformations, the number of validated semantics-preserving transformations, and the contribution of these transformations to fault detection.

\item \textbf{RQ6: How does \sys{} compare with existing testing approaches?}
This RQ compares \sys{} with prior testing tools in terms of effectiveness and precision, including the number of generated mutants, inconsistency-triggering mutants, detected unique faults, and testing false positives.
\end{itemize}

\noindent \textbf{Experimental Setup.}
We evaluated \sys{} on the six Rust SAST tools introduced in Section~\ref{sec:study}.
These tools cover a range of analysis paradigms, including rule-based matching, data-flow analysis, symbolic reasoning, and concurrency-oriented checking, allowing us to assess whether \sys{} is effective across tools with different internal designs.
As described in Section~\ref{sec:implementation}, \sys{} uses input programs collected from the official test suites of Rust SAST tools as the initial seed corpus.
We use OpenAI GPT-5.3 to synthesize program transformations and iteratively repair implementation errors in the generated transformations.
Each candidate transformation is then validated using short executable programs from Code4Bench~\cite{code4bench} by checking whether the transformed program preserves the same input-output behavior as the original.
Only validated transformations are used for fault detection.
Prior studies have shown that official test suites are effective seed corpora for SAST tools~\cite{statfier, annatester}; accordingly, we use them in our study.
For each seed test case, we reuse the SAST configuration provided by the original tool test suite when available; otherwise, we adopt the default recommended configuration.
All experiments were conducted on a server equipped with an Intel(R) Xeon(R) CPU Max 9468 processor, 2 TB of memory, and running Ubuntu 24.04.

\subsection{RQ4: Assessing Unique Fault Detection Capability}
\label{sec:rq4}
In this RQ, we evaluate the number of unique faults detected by \sys{} in Rust SAST tools.
For each potential fault reported by \sys{}, we manually inspect the seed--mutant pair and the corresponding analysis reports to determine whether it is a true SAST tool fault.

Table~\ref{tab:rq4-fault-detection} summarizes the faults identified by \sys{} in six Rust SAST tools.
For each tool, we report the total number of fault candidates and their type distribution, including false positives, false negatives, and crashes.
Overall, \sys{} identified \faultcnt{} unique faults across the six tools, all of which have been reported to the developers, and \confirmedfaultcnt{} have been confirmed by the developers.
These results demonstrate that \sys{} is effective in uncovering faults in Rust SAST tools.
In addition, \sys{} generates 88,453 mutant programs, 2,922 of which are fault-triggering, including 282 FP mutants, 2,586 FN mutants, and 54 crash mutants.

\begin{table}[tb]
\caption{Distribution of Detected Faults Across SAST Tools}
\centering
\footnotesize
\setlength{\tabcolsep}{2pt}
\begin{tabular}{l|r|r|r|r|r|r|r}
\toprule
\textbf{Tool} & \textbf{Semgrep} & \textbf{Prusti} & \textbf{LockBud} & \textbf{MirChecker} & \textbf{MIRAI} & \textbf{Rudra} & \textbf{Total} \\
\hline
\# FP & 3 & 6 & 2 & 2 & 2 & 0 & 15 \\
\# FN & 17 & 3 & 3 & 1 & 2 & 1 & 27 \\
\# Crash & 0 & 3 & 0 & 1 & 0 & 0 & 4 \\
\hline
\textbf{Total} & 20 & 12 & 5 & 4 & 4 & 1 & \faultcnt{} \\
\bottomrule
\end{tabular}
\label{tab:rq4-fault-detection}
\end{table}

\begin{findingbox}
\sys{} demonstrates strong unique fault detection capability for Rust SAST tools: it uncovered \faultcnt{} unique faults across all six evaluated tools, including 15 false positives, 27 false negatives, and 4 crashes, of which \confirmedfaultcnt{} have been confirmed by developers.
\end{findingbox}

\subsection{RQ5: Assessing Quality of Generated Transformations}
\label{sec:rq5}
In this RQ, we evaluate the transformation generation capability of \sys{}.
Specifically, we investigate three questions:
(1) how many transformations can be generated from Rust language references,
(2) how many of them are invalid and must be filtered out, and
(3) how useful the remaining valid transformations are for exposing faults of SAST tools.

\noindent \textbf{Transformation Quality.}
We first measure the number of candidate transformations produced by the LLM-assisted generation process.
These transformations are then first validated using the executable equivalence procedure described in Section~\ref{sec:validation}.
We consider a transformation invalid if it appears plausible at the specification level but either fails to preserve program behavior or produces uncompilable mutants during validation.
After querying the LLM with the language reference, \sys{} generated 163 program transformations.
The validation mechanism filtered out 132 of them, retaining 31 valid semantics-preserving transformations.
This result underscores the importance of the validation step: without behavioral checking, unsound transformations could be incorrectly attributed to faults in the target tools, while duplicate or trivial transformations could consume testing resources unnecessarily and reduce the overall efficiency of the testing process.
We further manually inspected the remaining transformations, removing duplicates and excluding those that introduced only minor syntactic variations, such as mutating \verb|42u32| into \verb|0x42u32| or transforming the string \verb|r#"foo"bar\baz"#| into \verb|"foo\"bar\\baz"|.
These simple transformations provide little benefit for generating mutant programs.
Totally, \sys{} synthesized 17 valid and non-trivial program transformations.
Table~\ref{tab:transformation} provides a detailed illustration of all synthesized program transformations.
Each row represents a transformation, including its index, description, and code example.
We further analyze why some candidate transformations fail the semantics-equivalence check.
Twelve are due to misunderstandings or misuse of closures.
Nine result from syntactically invalid transformations.
Four are caused by unintended changes to scope or name binding.
The remaining three are respectively caused by improper handling of Rust ownership, a misunderstanding of Rust attribute semantics, and a misunderstanding of Rust expression evaluation semantics.

\begin{table*}[htbp]
\caption{Detailed Illustration of Synthesized Program Transformations}
\centering
\footnotesize
\setlength{\tabcolsep}{2pt}
\begin{tabular}{p{0.17\textwidth} p{0.31\textwidth} p{0.17\textwidth} p{0.31\textwidth}}
\toprule
\textbf{Transformation} & \textbf{Code Example} &
\textbf{Transformation} & \textbf{Code Example} \\
\midrule

\makecell[l]{T1: Replace format!\\with fmt::format} &
\begin{minipage}[t]{\linewidth}
\vspace{-1.2\baselineskip}
\begin{lstlisting}[style=mystyle, numbers=none, frame=none]
- format!(args);
+ format(format_args!(args));
\end{lstlisting}
\end{minipage}
&
\makecell[l]{T2: Replace panic!\\with panic\_any} &
\begin{minipage}[t]{\linewidth}
\vspace{-1.2\baselineskip}
\begin{lstlisting}[style=mystyle, numbers=none, frame=none]
- panic!("message");
+ panic_any("message");
\end{lstlisting}
\end{minipage}
\\
\midrule

T3: Add explicit return &
\begin{minipage}[t]{\linewidth}
\vspace{-1.2\baselineskip}
\begin{lstlisting}[style=mystyle, numbers=none, frame=none]
- fn f() -> T { expr }
+ fn f() -> T { return expr; }
\end{lstlisting}
\end{minipage}
&
\makecell[l]{T4: Wrap function body\\in closure} &
\begin{minipage}[t]{\linewidth}
\vspace{-1.2\baselineskip}
\begin{lstlisting}[style=mystyle, numbers=none, frame=none]
- fn name(...) { body }
+ fn name(...) { (||{ body })() }
\end{lstlisting}
\end{minipage}
\\
\midrule

\makecell[l]{T5: Replace vec!\\with explicit construction} &
\begin{minipage}[t]{\linewidth}
\vspace{-1.2\baselineskip}
\begin{lstlisting}[style=mystyle, numbers=none, frame=none]
- let v = vec![1, 2, 3];
+ let v = [1, 2, 3].to_vec();
\end{lstlisting}
\end{minipage}
&
\makecell[l]{T6: Replace range literal\\with explicit construction} &
\begin{minipage}[t]{\linewidth}
\vspace{-1.2\baselineskip}
\begin{lstlisting}[style=mystyle, numbers=none, frame=none]
- start ..= end
+ std::ops::RangeInclusive::
+   new(start, end)
\end{lstlisting}
\end{minipage}
\\
\midrule

T7: Add an assignment &
\begin{minipage}[t]{\linewidth}
\vspace{-1.2\baselineskip}
\begin{lstlisting}[style=mystyle, numbers=none, frame=none]
- let var = expr; ... var ...
+ let var = expr;
+ let _new_var = var; 
+ ... _new_var ...
\end{lstlisting}
\end{minipage}
&
\makecell[l]{T8: Add explicit lifetime\\subtyping annotation} &
\begin{minipage}[t]{\linewidth}
\vspace{-1.2\baselineskip}
\begin{lstlisting}[style=mystyle, numbers=none, frame=none]
- let x: &'a T = expr;
+ let x: &'a T = {
+   let _temp: &'static T = expr; 
+   _temp };
\end{lstlisting}
\end{minipage}
\\
\midrule

\makecell[l]{T9: Replace ?\\with explicit match} &
\begin{minipage}[t]{\linewidth}
\vspace{-1.2\baselineskip}
\begin{lstlisting}[style=mystyle, numbers=none, frame=none]
- let x = expr?;
+ let x = match expr { 
+  Ok(val)=>val,
+  Err(e)=>return Err(e.into()),
+ };
\end{lstlisting}
\end{minipage}
&
\makecell[l]{T10: Replace if let\\with match} &
\begin{minipage}[t]{\linewidth}
\vspace{-1.2\baselineskip}
\begin{lstlisting}[style=mystyle, numbers=none, frame=none]
- if let PATTERN=value {...}
- else {...}
+ match value {
+   PATTERN => {...},
+   _ => {...}, }
\end{lstlisting}
\end{minipage}
\\
\midrule

\makecell[l]{T11: Replace while\\with loop} &
\begin{minipage}[t]{\linewidth}
\vspace{-1.2\baselineskip}
\begin{lstlisting}[style=mystyle, numbers=none, frame=none]
- while cond { body }
+ loop {
+   if !(cond) { break; } 
+   body }
\end{lstlisting}
\end{minipage}
&
\makecell[l]{T12: Add unreachable\\match arm} &
\begin{minipage}[t]{\linewidth}
\vspace{-1.2\baselineskip}
\begin{lstlisting}[style=mystyle, numbers=none, frame=none]
- match expr { arms }
+ match expr { 
+  arms, 
+  _ if false => unreachable!()}
\end{lstlisting}
\end{minipage}
\\
\midrule

\makecell[l]{T13: Rewrite parameters\\as references} &
\begin{minipage}[t]{\linewidth}
\vspace{-1.2\baselineskip}
\begin{lstlisting}[style=mystyle, numbers=none, frame=none]
- fn func(param: Type)
-   -> RetType { body }
+ fn func(param: &Type)
+   -> RetType {
+   let param = *param; body }
\end{lstlisting}
\end{minipage}
&
\makecell[l]{T14: Rewrite parameters\\as RefCell references} &
\begin{minipage}[t]{\linewidth}
\vspace{-1.2\baselineskip}
\begin{lstlisting}[style=mystyle, numbers=none, frame=none]
- fn func(param: Type) 
-   -> RetType { body }
+ fn func(param: RefCell<&Type>)
+   -> RetType {
+   let param = **param.borrow(); 
+   body }
\end{lstlisting}
\end{minipage}
\\
\midrule

\makecell[l]{T15: Replace impl Trait \\ with generic parameter} &
\begin{minipage}[t]{\linewidth}
\vspace{-1.2\baselineskip}
\begin{lstlisting}[style=mystyle, numbers=none, frame=none]
- fn f(arg: impl Trait) { ... }
+ fn f<T: Trait>(arg: T) { ... }
\end{lstlisting}
\end{minipage}
&
\makecell[l]{T16: Add redundant\\let-else} &
\begin{minipage}[t]{\linewidth}
\vspace{-1.2\baselineskip}
\begin{lstlisting}[style=mystyle, numbers=none, frame=none]
- let pat = expr;
+ let pat = expr else {panic!()};
\end{lstlisting}
\end{minipage}
\\
\midrule

\makecell[l]{T17: Add higher-ranked\\lifetime substitution} &
\begin{minipage}[t]{\linewidth}
\vspace{-1.2\baselineskip}
\begin{lstlisting}[style=mystyle, numbers=none, frame=none]
- let f: &(fn(&'static T)->&'static T) = g;
+ let f: &(fn(&'static T)->&'static T) = { 
+   let _temp: &(for<'a> fn(&'a T)->&'a T) = g; _temp };
\end{lstlisting}
\end{minipage}
\\


\bottomrule
\end{tabular}
\label{tab:transformation}
\end{table*}

\begin{figure}[htbp]
\centering
\includegraphics[width=0.9\linewidth]{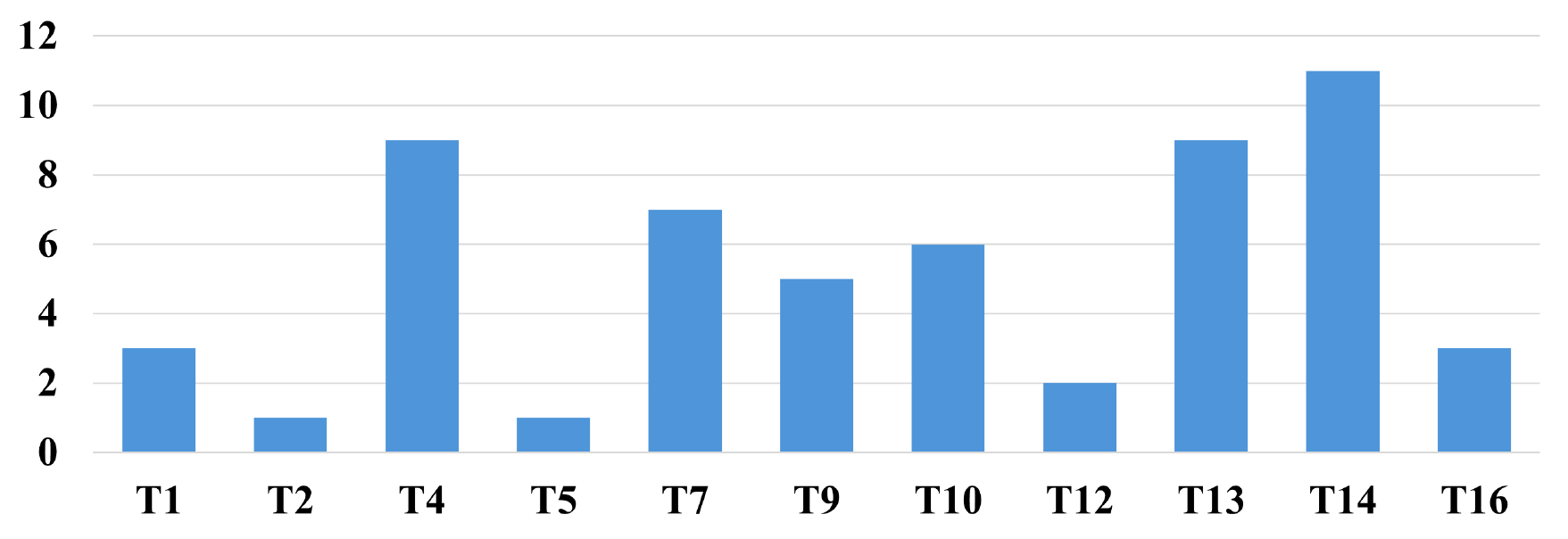}
\caption{Number of Detected Faults by each Transformation}
\label{fig:detected_faults}
\end{figure}

\noindent \textbf{Fault Detection Capability of Each Transformation.}
Next, we examine the number of faults exposed by each valid program transformation.
This analysis helps us understand which kinds of semantics-preserving transformations are most effective for testing Rust SAST tools.
Figure~\ref{fig:detected_faults} presents the number of faults discovered by different transformations.
Note that the transformation index (Tx) is consistent with Table~\ref{tab:transformation}.
The results indicate that the fault-finding ability of transformations is uneven.
Some transformations, such as T4, T13, and T14, reveal substantially more faults than others.
Notably, the three program transformations that exposed the most faults are all centered on Rust-specific language constructs.
T4 wraps a function body in an immediately invoked closure, exercising the closure and block-expression semantics of Rust.
T13 rewrites by-value parameters as references, explicitly taking the ownership, borrowing, and dereferencing mechanisms of Rust into account.
T14 further leverages \texttt{RefCell}, a Rust-specific interior-mutability abstraction. These results suggest that effective testing of Rust SAST tools benefits from Rust-aware transformations that are designed around the language’s core constructs and semantic constraints, rather than from generic syntax-level rewrites alone.
However, six transformations still do not reveal any faults.
Overall, 65\% of the synthesized and validated transformations are effective in exposing faults in Rust SAST tools.

\begin{findingbox}
\sys{} synthesizes 17 valid and non-trivial semantics-preserving transformations; 65\% of them expose at least one unique fault, showing that reference-guided transformation generation is practical for testing Rust SAST tools.
\end{findingbox}

\subsection{RQ6: Comparison with Existing Testing Approaches}
\label{sec:rq6}
In this research question, we compare \sys{} with \statfier{}~\cite{statfier}, a state-of-the-art testing approach for SAST tools.
For this comparison, we use four evaluation metrics: the number of generated mutants, the number of inconsistency-triggering mutants, the number of unique confirmed faults, and the number of false positives.
Table~\ref{tab:comparison} presents the comparison results.
Compared with \statfier{}, \sys{} demonstrates substantially stronger testing capability: it generates more buggy mutants and discovers more unique faults.
Moreover, \sys{} introduces no testing false positives, whereas \statfier{} produces six, indicating that our approach achieves both higher effectiveness and better precision.
Overall, \sys{} achieves better fault-detection performance than \statfier{} because it leverages two complementary advantages.
First, its transformations are aware of language features, enabling \sys{} to specifically target Rust constructs that frequently lead to analysis failures.
Second, its transformation space is extensible, as new transformations can be derived from evolving language reference documents rather than being limited to a fixed handcrafted rule set, as in \statfier{}.

\begin{table}[htbp]
\caption{Comparison with the SOTA Testing Approach}
\label{tab:comparison}
\centering
\footnotesize
\setlength{\tabcolsep}{2pt}
\begin{tabular}{l|r|r|r|r}
\toprule
Tool & \# Mutant & \# Buggy Mutant & \# Unique Fault & \# False Positive \\
\hline
\sys{} & 88453 & 2922 & \faultcnt{} & 0 \\
\hline
\statfier{} & 38140 & 1032 & 12 & 6 \\
\bottomrule
\end{tabular}
\end{table}

\begin{findingbox}
Compared with the SOTA testing approach, \sys{} achieves better fault-finding capability, while remaining extensible to new Rust language features through reference-guided transformation synthesis.
\end{findingbox}

\section{Lessons and Implications}
\label{sec:lessons}

Our empirical study yields several \textbf{\emph{actionable lessons}} for Rust SAST tool industrial developers and academic researchers.
These lessons not only explain the current limitations of Rust SAST in practice, but also directly motivate the design of our testing framework.

\noindent \textbf{Lesson 1: Real-world Rust vulnerability patterns should guide SAST coverage.}
From RQ1 and RQ2, we observe that the six studied tools cover only a limited set of vulnerability types, and their practical effectiveness is still unsatisfactory.
Although some tools support a relatively broad set of CWE IDs, their coverage only partially overlaps with the vulnerabilities that actually appear in real-world Rust applications.
Moreover, most of tools exhibit high FN rates and suffer from high FP rates.
These results suggest that Rust SAST development should not only expand checker coverage, but should prioritize vulnerability patterns that are frequent or security-critical in real Rust projects, such as concurrency bugs, memory-access violations, use-after-free patterns around unsafe code, and resource-management issues.
For practitioners, these results also imply that current Rust SAST outputs should be treated as auxiliary security evidence rather than as a standalone assurance mechanism, and should be combined with code review, fuzzing, dependency auditing, and manual inspection of high-risk code.

\noindent \textbf{Lesson 2: Rust-specific features require explicit modeling in SAST analysis.}
Our FN study reveals that a substantial number of missed vulnerabilities are not caused by a lack of checkers alone, but by unexpected analysis failures, including crashes, hangs, internal compiler errors, and toolchain incompatibilities.
This finding highlights an often overlooked issue in Rust SAST tools: \emph{even a tool with theoretically strong analysis capability is of limited practical value if it cannot stably analyze modern Rust code}.
Therefore, SAST developers should continuously test their tools against evolving Rust compiler versions, Cargo features, essential third-party crates, and emerging language idioms.
Similarly, SAST benchmarks should report robustness failures explicitly rather than filtering them out as engineering issues.
Improving Rust SAST therefore requires not only better vulnerability detectors, but also stronger engineering support for robustness, version compatibility, graceful failure handling, and long-term maintainability.

\noindent \textbf{Lesson 3: Evolution-aware SAST testing requires extensible test generation.}
Our study shows that many failures are closely tied to evolving language constructs and ecosystem-specific idioms.
Consequently, manually curated testing patterns can quickly become stale.
Testing frameworks should therefore support continuous extension as new language features, APIs, and programming idioms emerge.
For semantics-preserving transformations, this means making the transformation space derivable from the Rust Reference and ecosystem idioms, enabling it to be systematically updated as the language and ecosystem evolve, rather than being confined to handcrafted rewrites.
This insight motivates the design of \sys{}, so that the transformation set can be extended as the language and ecosystem evolve, rather than being confined to predefined handcrafted rules.

\noindent \textbf{Lesson 4: Rust-specific language features require explicit support in SAST tools.}
A key finding of RQ3 is that many analysis failures are rooted in misunderstanding Rust-specific language features, such as language constructs, and ownership propagation.
This lesson is particularly important as Rust security analysis cannot rely only on generic static analysis techniques inherited from other languages.
They should also account for Rust-specific idioms such as iterator and combinator chains, trait-based concurrency guarantees, and standard library provided abstractions such as \texttt{Mutex}, \texttt{RwLock}, and their ecosystem variants.
Without such feature modeling, both precision and recall will remain fundamentally limited.

\section{Threats to Validity}
\noindent\textbf{External.}
Our findings may not generalize beyond the evaluated SAST tools, Rust, or the implemented transformations.
Although many other SAST tools exist (such as PMD~\cite{pmd} and Error Prone~\cite{sadowski2018lessons}), our evaluation covers only \numtool{} popular tools and use only Rust programs as inputs, even though some tools support multiple languages (e.g., \semgrep{} also analyzes C/C++). 
To mitigate this threat, we include transformations shown to be effective in prior work, and our experiments indicate that they are general enough to uncover bugs in the evaluated analyzers. 
For root-cause categorization, two authors independently inspected analysis reports and execution logs for each fault and resolved disagreements through discussion.
We further reported identified faults to developers by filing \faultcnt{} bug reports.

\noindent\textbf{Internal.}
Our code and automated scripts may have bugs that can affect the results.
To mitigate this threat, we tested the scripts on representative cases, manually inspected intermediate results, and made our implementations publicly available for inspection and replication.

\section{Related Work}
\label{sec:related_work}

\noindent \textbf{Testing SAST Tools.}
Several techniques have been proposed for testing SAST tools~\cite{2022icpcfinder, klinger2019differentially, ase2024minotaur, ase2024sherlock, fse2024sabugs}.
Some prior work employs differential testing to address the oracle problem.
However, such approaches typically rely on common components shared by two SAST tools, such as bug detectors~\cite{2022icpcfinder} and program-representation construction algorithms~\cite{sascope}, which limits their scalability.
Other studies adopt specification-based testing~\cite{cgo2020testsa}, which requires the correct behavior of an analyzer to be manually and precisely formalized, making it difficult and often impractical in practice.
To mitigate the oracle problem, some work uses equivalence relations as metamorphic relations for testing SAST tools~\cite{statfier, annatester}.
Similarly, we leverage equivalence relations by generating mutants and checking whether equivalent mutants yield inconsistent analysis results.
Unlike prior work, however, we further exploit LLMs to automatically derive and implement new semantics-preserving program transformations, rather than relying on a fixed set of transformations.

\noindent \textbf{Studies on SAST Tools.}
As a widely adopted technique in software security, SAST has been extensively studied in prior work~\cite{sast1, sast3, sast4}.
Existing studies have primarily focused on evaluating the performance and effectiveness of SAST tools~\cite{sast1}, examining how developers use them in real-world settings~\cite{sast4}, and improving their detection capabilities through tool-combination strategies~\cite{sast3}.
However, none of this work has explored the use of LLMs to support the testing of SAST tools by identifying their faults.
Our work fills this gap by leveraging LLMs to synthesize semantics-preserving program transformations to uncover weaknesses in SAST tools and provide targeted insights for improvement, thereby enhancing their precision and reliability in practical security-testing workflows.

\noindent \textbf{LLM-based Testing Approaches.}
LLM-based techniques have been explored for testing a variety of system software.
However, to the best of our knowledge, no prior work has investigated the use of LLMs for testing SAST tools.
Among existing studies, LLM-based compiler testing approaches~\cite{xia2024fuzz4all} are the most closely related to our work.
Nevertheless, their methodologies are not directly applicable to SAST tools for two main reasons.
First, their test oracles cannot be directly transferred because the fault manifestations of compilers and SAST tools differ substantially, especially for logic faults.
For example, Fuzz4All~\cite{xia2024fuzz4all} focuses only on crash bugs, making its oracle design relatively straightforward.
By contrast, our work targets SAST tools, whose faults are often reflected in incorrect or inconsistent analysis results rather than crashes, thus requiring fundamentally different oracle designs.
Second, the program transformations used in compiler testing are insufficient for comprehensively evaluating SAST tools, e.g., Fuzz4All relies on LLMs to directly perform program transformations, making the process less predictable.
In contrast, our approach automatically derives semantics-preserving transformations from language references and implements them explicitly, avoiding direct reliance on LLMs for transformation.
This design improves transparency and reduces the risk of hallucinations during mutant generation.
Moreover, transformations developed for other domains are often not directly applicable to SAST tools because of differences in programming language grammar and analysis objectives.
\section{Conclusion}
We presented the first systematic empirical study of Rust SAST tools, examining their capability, effectiveness, and failure causes.
Across six representative tools, we find high FP and FN rates and limited usability and reliability. Many failures arise from mishandling Rust-specific semantics and from the rapidly evolving Rust ecosystem.
These findings motivate \sys{}, a framework for robustness testing of Rust SAST tools via LLM-assisted, reference-guided synthesis of semantics-preserving transformations.
\sys{} generates language feature aware transformations from Rust language reference, validates them by executable equivalence checking, and detects faults.
Evaluation has shown that \sys{} exposes \faultcnt{} unique faults in multiple Rust SAST tools and outperforms existing baselines.

\bibliographystyle{IEEE-Reference-Format}
\bibliography{reference}

\end{document}